\documentclass[aps,pra,twocolumn,floatfix]{revtex4-1}
\usepackage{graphicx}
\usepackage{dcolumn}
\usepackage{bm}
\usepackage{multirow}
\usepackage{ulem}
\usepackage{color}
\usepackage{longtable}
\usepackage{subfigure}
\usepackage{amssymb}
\usepackage{amsmath}
\usepackage{upgreek}
\usepackage{latexsym,epsfig}
\usepackage{float}
\usepackage[figuresright]{rotating}
\usepackage{upgreek}
\usepackage{ulem}
\usepackage{booktabs}

\begin{document}
	\title{Investigating properties of heavy and superheavy atomic systems with $p^{3}$ configurations}
	
	\author{H. X. Liu$^{1,2}$} 
	\author{Y. M. Yu$^{2,4}$} \email{Email: ymyu@aphy.iphy.ac.cn} 
	\author{B. B. Suo$^{3}$} \email{Email: bsuo@nwu.edu.cn} 
	\author{Y. Liu$^1$}      \email{Email: yongliu@ysu.edu.cn}
	\author{B. K. Sahoo$^5$} \email{Email: bijaya@prl.res.in}
	
	\affiliation{$^1$State Key Laboratory of Metastable Materials Science and Technology \& Key Laboratory for Microstructural Material Physics of Hebei Province, School of Science, Yanshan University, Qinhuangdao, 066004, China}
	\affiliation{$^2$Beijing National Laboratory for Condensed Matter Physics, Institute of Physics, Chinese Academy of Sciences, Beijing 100190, China}
	\affiliation{$^3$Institute of Modern Physics, Northwest University, Xi’an, Shanxi 710069, China}
	\affiliation{$^4$University of Chinese Academy of Sciences, 100049 Beijing, China}
	\affiliation{$^5$Atomic, Molecular and Optical Physics Division, Physical Research Laboratory, Navrangpura, Ahmedabad 380009, India}
	
	\begin{abstract}
		We have investigated energies and spectroscopic properties such as lifetimes, $g_J$ factors, and hyperfine structure constants of the neutral atoms P
		through Mc belonging to Group-15, singly ionized atoms S$^+$ through Lv$^+$ of Group-16 and doubly ionized atoms Cl$^{2+}$ through Ts$^{2+}$ of Group-17 of
		the periodic table. These elements have $np^{3}$ configurations with $n=3-7$, which are highly open-shell and expected to exhibit strong 
		electron correlation effects. We have used four-component Dirac-Coulomb Hamiltonian along with Gaunt term and a relativistic effective core potential 
		through the relativistic multi-reference configuration interaction method to perform the calculations with sufficient accuracy and compare the results 
		with the available literature data. These comparisons suggest that our predicted values, for which experimental data are not available, are reliable 
		enough to be useful for future applications.
	\end{abstract}
	
	\date{\today}
	
	\maketitle
	
	\section{Introduction}
	
	There have been growing interest to study atomic spectroscopy of ions with $np^3$ configurations due to their demands to use them in the 
	astrophysical analysis, diagnostic tools for tokamak, laboratory and high-temperature-fusion plasma etc. to name a few \cite{Biemont-MNRAS-1983,
		Roederer-APJL-2012,Zhang-A&A-2013,Quinet-CJP-2017,Trabert-atoms-2020}. Generally, transition lines among fine-structure splitting are found 
	in the far ultraviolet region. However, in the heavy neutral atoms and ions with less degree of ionization such as singly and doubly ionized 
	systems one can observe these lines in the visible region and accessible by the lasers to carry out high-precision measurements using the 
	corresponding atomic systems. For example, the magnetic dipole (M1) transition within the fine-structure splitting of the $np^3$ configuration 
	in the Bi atom were considered for measuring parity nonconserving electric dipole amplitude \cite{Dzuba-EL-1988,Macpherson-PRL-1991}. These 
	transitions from various atomic species are also being considered for making optical clocks \cite{Yudin-PRL-2014, Yu-PRA-2018, Beloy-PRL-2020, 
		Allehabi-PRA-2022}.
	
	Bi\'{e}mont, et al. have conducted a systematic investigation of atomic energy levels comprising $np^k$ configurations, $k=1-5$, with the aim of 
	interpreting, identifying, and analyzing many lines relevance for astrophysical and high-temperature plasmas studies \cite{Biemont-PS-1986,
		Biemont-AASS-1995,Biemont-PS-1996,Biemont-PS-1986-1,Biemont-PS-1987}. These calculations, however, were mainly based on the approximate relativistic 
	Hartree-Fock (HFR) method \cite{Cowan}. The $np$ and $np^5$ configurations, containing one particle or hole in the $p$ shells, and the $np^2$ and 
	$np^4$ configurations, containing two particles or holes in the $p$ shells, are comparatively simpler to handle in the many-body approach than the 
	systems having the $np^3$ configurations owing to the fact that there are three valence electrons present in the $p$ shell which can give rise 
	a large number of strongly interacting configuration interaction space for the determination of atomic wave functions. Despite the increasing availability of 
	computing power in recent years, rigorous calculations of properties of atomic systems with $np^3$ configurations still appear to be lacking. Among 
	the limited studies, Rynkun, et al. have calculated energy levels and radiation properties of the systems with $3p^3$ configurations for 
	S$^+$, Cl$^{2+}$, and Ar$^{3+}$ within the P-isoelectronic sequences by using multi-configuration Dirac-Hartree-Fock (MCDHF) method 
	\cite{Rynkun-AA-2019}, Skripnikov, et al. have calculated the electronic structure of Bi by using the relativistic coupled-cluster method 
	\cite{Skripnikov-PRC-2021} and Hussein has conducted the calculations on the neutral phosphorus and two of its ions with $3s^23p^n$ configurations
	using the configuration interaction (CI) method \cite{Hussein-PRA-2022}. 
	
	Besides the considered $p^3$ systems are being highly open-shell with three valence electrons, they also have a large number of occupied electrons. 
	Thus, determining wave functions of these systems would require multi-configuration treatment. It would really be a challenge to deal with a 
	multi-reference many-body theory with such large number of electrons. One way to address this challenge is to utilize adopt a hybrid method. It 
	means that all the ccupied electrons of the system can be classified into two groups -- core and valence electrons. Then, correlations due to 
	the valence electrons can be treated more rigorously than the closed-core orbitals. The CI method is more suitable to include correlation
	effects due to the valence electrons while less computational expensive method like many-body perturbation theory (MBPT method) can be employed to 
	take into account correlations among the core core electrons and core-valence electron interactions. Such a combined CI$+$MBPT method has been 
	used by several groups to study atomic properties of a wide range of atomic systems \cite{Dzuba-PRA-1996,Dzuba-PRA-2017,Berengut-PRA-2016,
		Savukov-PRA-2020}. Later this hybrid approach has been extended to CI+all-order method that treats core and core-valence correlations using the 
	coupled-cluster method \cite{Safronova-PRA-2018}. 
	
	On the other hand, all possible correlations arising due to both the core and valence electrons can be treated on equal footing using the CI method. 
	To allow participation more number of electrons for producing a large configuration space in the multi-reference CI (MRCI) method, the general active 
	space (GAS) technique has been developed. It offers a more comprehensive treatment to the the multivalent configurations in the highly open-shell systems 
	\cite{Olsen-JCP-1988,Fleig-JCP-2001,Fleig-JCP-2003,Fleig-JCP-2006}. In this approach, the active orbitals are divided into subspaces with flexible 
	restrictions on the allowed electronic occupation and consider excitations in each subspace, which is tailor made for accurate study of atomic 
	properties of heavier open-shell systems. For this reason, the performance of the MRCI method has been found to be amazing that produces 
	fine-structure splitting of many one-valent $p$-, $d$-, and $f$-block atomic systems closed to the  experimental values \cite{Hu-JCTC-2020}.
	
	In this study, we investigate the energies and spectroscopic properties of the neutral P through Mc, the singly ionized S$^{+}$ through Lv$^{+}$,
	and the doubly ionized Cl$^{2+}$ through Ts$^{2+}$ atoms from Groups 15-17 of the periodic table having $np^{3}$ configurations ($n=3-7$) using the
	MRCI method. The complete basis set (CBS) limits of the energies are used to estimate uncertainties in the calculations. First, we study the properties 
	using a relativistic effective core potential (RECP) in the atomic Hamiltonian and then considering the Dirac-Coulomb (DC) Hamiltonian with the Gaunt term.
	We have also analyzed results with different combinations of basis functions to test reliability of the calculations. Our reported properties include
	lifetimes, Land\'{e} $g_J$ factors, and magnetic dipole ($A$) and electric quadrupole ($B$) hyperfine structure constants of the 
	fine-structure partner states with $np^{3}$ configurations. We find that the forbidden transitions among these fine-structure partner states are in the 
	optical region that can be used for carrying out precision measurements and other reported properties can be useful in the analysis of astrophysical and 
	plasma diagnostic processes. 
	
	\section{Methods and technical details}
	
	We discuss here the methods and various procedures adopted to carry out the calculations and estimating uncertainties to the investigated 
	properties. All the ingrediants used in the computations are available with the recently eleased version of the Dirac program suite 
	\cite{Dirac,Dirac2020}. This program is well known in the community, however for the sake of completeness we outline below some the important 
	points for better understanding of the accuracy of the calculated quantities.
	
	\subsection{The DC Hamiltonian}
	
	The DC Hamiltonian in a.u. is given by
	\begin{eqnarray}
		\hat{H}&=&\sum_i [c(\vec { \bm{\alpha}}\cdot \vec {\bf p})_i+(\bm \beta-1)_i m_0c^2+V_{nuc}(r_i) ] + \nonumber \\
		&& \sum_{i<j}\bigg[\frac{1}{r_{ij}}-\frac{1}{2}\frac{ {\vec {\bm \alpha}}_i \cdot {\vec { \bm \alpha}}_j }{r_{ij}}\bigg], \label{DCG}
	\end{eqnarray}
	where $ \vec {\bm \alpha}$ and  $\bm \beta$ are the Dirac matrices, $\vec {\bf p}$ is the momentum operator, $m_0c^2$ denotes the rest mass energy
	of an electron with the speed of light $c$ and $V_{nuc}(r)$ is the nuclear potential. The last term in Eq. (\ref{DCG}) represents the Gaunt interaction,
	which is the leading term of the Breit interaction.
	
	First, we obtained the Dirac-Hartree-Fock (DHF) wave function for the above Hamiltonian using the correlation consistent basis sets, dyall.aaeXz, 
	with $X$ = 2, 3, and 4 denoting for the double, triple, and quadruple-$\xi$ basis sets that are developed by Dyall and co-workers 
	\cite{Visscher-JCP-1996, Dyall-TCA-1998,Dyall-TCA-2002,Dyall-TCA-2006,Dyall-TCA-2012-1,Dyall-TCA-2012-2,Dyall-TCA-2016}. These basis sets are 
	optimized to account for correlations among electrons from all core and valence shells and polarization of the $d$ shells. They are also augmented 
	with diffused and tight functions as designed for the p-group elements to ensure high accuracy in the calculations. For the elements with $3p$ and 
	$4p$ valence orbitals, the non-relativistic correlation-consistent aug-cc-pVXZ basis sets, as developed by Dunning and co-workers 
	\cite{Woon-JCP-1993,Wilson-JCP-1999,DeYonker-JCP-2007}, are also used in this study. These basis sets are particularly appropriate for the light 
	elements and include larger configuration space, up to 7 $\xi$ for the $3p$ elements and 5 $\xi$ for the $4p$ elements. Such larger 
	size of basis function is advantageous for achieving high precision in the CBS limit.

	\subsection{The RECP approach}
	
	The RECP method is an efficient for heavier atomic systems with reasonable accuracy at the less computational 
	cost. It treats the inner core electrons and nuclei as chemically inert entities, and includes the remaining electrons (outer core and
	valence electrons) explicitly in the Hamiltonian. It accounts for the dominant relativistic effects through the nonrelativistic formulation. 
	The effective Hamiltonian in this method is given by \cite{Stoll-JCC-2002,Lim-JCP-2005,Dolg-MMAQC-2000}
	\begin{equation}
		H_v=\sum_{i=1}^{n_v} \bigg(-\frac{1}{2}\Delta_i+V_i^{SOPP}\bigg)+\sum_{i < j}^{n_v} \frac{1}{r_{ij}},
	\end{equation}
	where $n_v=Z-Q_c$ is the number of outer correlated electrons for atomic number $Z$ and charge of the ionic core $Q_c$, and the $\Delta$ operator is given by
	\begin{eqnarray}
		\Delta_i = {\vec \sigma}_i \cdot  {\vec p}_i \  \frac{2 c^2}{E_i -V_{nuc}(r_i) + 2c^2} \ {\vec \sigma}_i \cdot {\vec p}_i,
	\end{eqnarray}
	with the Pauli spinor $\sigma$, momentum operator $p$ and $E_i = c \sqrt{p_i^2 + c^2}$. In the above expression, $V_i^{SOPP}$ is the pseudopotential (PP)
	seen by the outer actively correlated electrons due to the presence of the atomic core. Kim, et al, have implemented two-component spin-orbit relativistic 
	effective core potential (SOREP) methods in the DIRAC package and is given by \cite{Kim-BKCS-2013}
	\begin{eqnarray}
		V_i^{SOPP}&=&-\frac{Q_c}{r_{i}}+U^{AREP}+U^{SO}. 
	\end{eqnarray}
	Here, the first term represents the Coulomb interactions between the $i$-th and the atomic core, the second term is the spin-average part of the
	PP and the third term takes into account spin-orbit interactions. Explicitly, these potential terms are given by
	\begin{eqnarray}
		U^{AREP}&=&U^{AREP}_L(r_i)+ \nonumber \\
		&&\sum^{L-1}_{l=0}\sum^l_{m=-l}[U^{AREP}_l(r_i)-U^{AREP}_L(r_i)]| lm \rangle \langle lm |\nonumber \\
	\end{eqnarray}
	such that total angular momentum $j$ is avaraged out to express in terms of orbital quantum numbers $l$ and $m$, and the spin-orbit interaction potential is
	given by
	\begin{eqnarray}
		U^{SO}&=&s\cdot\sum^{L}_{l=1}\frac{2}{2l+1}\Delta U^{SOREP}_l(r_i) \nonumber \\
		&&\sum^l_{m=-l}\sum^l_{m'=-l}|lm\rangle\langle lm|l|lm'\rangle \langle lm'|
	\end{eqnarray}		
	with
	\begin{equation}
		U^{AREP}_l(r_i)=\sum_kC_{lk}r_i^{n_{lk}-2}\exp({-\alpha_{lk}r_i^2})
	\end{equation}
	and 
	\begin{eqnarray}
		\Delta U^{SOREP}_l(r_i)=A_{lk}r_i^{n_{lk-2}}\exp({-\alpha_{lk}r_i^2})\bigg /\frac{2}{2l+1} .
	\end{eqnarray}
	In these definitions, $n_{lk}$, $\alpha_{lk}$, $C_{lk}$, and $A_{lk}$ are known as the PP parameters that are taken from the Stuttgart-Cologne
	PP library package. 
	
	In the present work, we consider both small-core  \cite{Metz-JCP-2000, Peterson-JCP-2003-1,Peterson-JPCA-2006} and large-core PPs 
	\cite{Stoll-JCP-2002} that are developed $P$-group elements. In details, the small-core PP ECPDS10MDFSO PPs are used for As, Se$^+$, and 
	Br$^{2+}$, where $Q_c=10$ represents the $(1s,2s,2p)$ core, while the $3s, 3p, 3d, 4s, 4p$ spinors are considered as the outer correlated orbitals;
	The small-core PP ECPDS28MDFSO PPs are used for Sb, Te$^+$, and I$^{2+}$, where $Q_c=28$ representing the $(1s-3d)$ core, while the $4s, 4p, 4d,
	5s, 5p$ spinors are taken as the outer active orbitals; The small-core PP ECPDS60MDFSO PPs are used for Bi, Po$^+$, and At$^{2+}$,  where $Q_c=60$ 
	meaning $(1s-4f)$ is the core and the remaining $5s, 5p, 5d, 6s, 6p$ are considered as the actively correlated orbitals. In the counter part, the 
	large-core PPs, ECPDS28MDFSO, ECPDS46MDFSO and ECPDS78MDFSO, are used for As to Br$^{2+}$, Sb to I$^{2+}$, and Bi to At$^{2+}$, respectively, where 
	$Q_c$ = 28, 46, and 78 represent $(1s-3d)$, $(1s-4d)$, and $(1s-5d)$ core, respectively. For the super heavy elements in the in the $7p$ row, we choose
	the ECPDS92MDFSO PP \cite{Hangele-JCP-2012,Hangele-JCP-2013}, where $Q_c$=92 represents the [Rn]$5f^{14}$ PP core, while the $6d, 7s, 7p$ shells are used 
	for accounting correlation effects.
	
	The RECP calculations are carried out with the help of PP-based correlation consistent basis sets, denoted as aug-cc-pVXZ-PP, 
	for $X$ = D, T, Q, and 5$\xi$ that are developed by Peterson, et al.. These basis functions are optimized to include the correlation of 
	electrons from the $ns$, $np$ and $(n-1)spd$ shells and are contracted adopting general schemes for the $p$-group elements
	\cite{Peterson-JCP-2003-1, Peterson-JCP-2003-2, Peterson-JCP-2010}.

	\begin{table*}[btp] 
		\caption{The general active space (GAS) model adopted in the MRCI calculation. The minimum (Min.) and maximum (Max.) numbers of the accumulated electron
			occupation, the number of Kramer pairs, the corresponding function type of the Kramer pairs, the allowed electronic excitation orders are given for per 
			GAS layer, where `S', `D', `T', and `Q' represent the singlets, doublets, triplets, and quadruplets of excitation. \label{tab:GAS}}
		{\setlength{\tabcolsep}{8pt} 
			\begin{tabular}{ccccccc}\hline\hline  \addlinespace[0.1cm]
				GAS layer No.	&	Min.	&	Max.	&	Number of Kramers pairs	&	Function type 	&	Allowed excitation levels	\\ \hline \addlinespace[0.1cm]
				\multicolumn{6}{l}{Core10SV2SDV3SDT}											\\
				I	&	9	&	10	&	5	&	$(n-1)d$, outer core	&	S	\\
				II	&	10	&	12	&	1	&	$ns$, outer core	&	S, D	\\
				III	&	12	&	15	&	3	&	$np$, valence	&	S, D, T	\\
				IV	&	15	&	15	&	m	&	rest, virtual	&		\\ [+1ex]
				\multicolumn{6}{l}{Core10SDV2SDTV3SDTQ} 											\\
				I	&	8	&	10	&	5	&	$(n-1)d$, outer core	&	S, D	\\
				II	&	9	&	12	&	1	&	$ns$, outer core	&	S, D, T	\\
				III	&	11	&	15	&	3	&	$np$, valence	&	S, D, T, Q	\\
				IV	&	15	&	15	&	m	&	rest, virtual	&		\\ [+1ex]
				\multicolumn{6}{l}{Core18SV2SDV3SDT}											\\
				I	&	17	&	18	&	9	&	$(n-1)spd$, outer core	&	S	\\
				II	&	18	&	20	&	1	&	$ns$, outer core	&	S, D	\\
				III	&	20	&	23	&	3	&	$np$, valence	&	S, D, T	\\
				IV	&	23	&	23	&	m	&	rest, virtual	&		\\ [+1ex]
				\multicolumn{6}{l}{V5SDTQ}											\\
				I	&	1	&	5	&	4	&	$nsp$, valence	&	S, D, T, Q	\\
				II	&	5	&	5	&	m	&	rest, virtual	&		\\ \hline \hline
		\end{tabular}}
	\end{table*}
	
	\subsection{Different GAS models in MRCI}
	
	After the DHF calculation using the DC Hamiltonian and RECP approximated Hamiltonian, we performed the MRCI calculation by using the 
	Kramers-restricted configuration interaction (KRCI) code \cite{Fleig-JCP-2001,Fleig-JCP-2003,Fleig-JCP-2006}. For the $np^3$ systems, we divided 
	the active spinors into three regimes, `outer core', `valence', and `virtual', apart form the inner core and high-lying orbitals that are frozen
	due to their less significant contributions to the electron correlation effects in the MRCI method. 
	
	Table \ref{tab:GAS} illustrates details about the GAS model that are used for the MRCI calculation for the considered $np^3$ systems. In the
	table, `Core10' means that the $(n-1)d$ outer core (or $(1s,2s,2p)$ outer core when $3p^3$) and `Core18' denotes the $(n-1)s,p,d$ outer core. 
	The `V2' and `V3' regions correspond to the $ns^2$ and $np^3$ electrons, respectively. The remaining unoccupied spinors consist of the virtual 
	subspace. Thus, by designing different types of the GAS model, we can investigate how the results converge with the increasing level of 
	excitations hierarchy due to electron correlation effects. For example, differences in the results between Core10SV2SDV3SDT and 
	Core18SV2SDV3SDT will reflect the variation in the results caused by the added inner-shelled electron excitations through `Core10' against 
	`Core18'. Similarly, differences in the results with the basis sets between Core10SV2SDV3SDT and Core10SDV2SDTV3SDTQ will indicate the 
	account of correlation effects included due to consideration of the higher level excitations when it increases from S to SD in layer I, from 
	SD to SDT in layer II type and from SDT to SDTQ in the layer III type. The `Core10SV2SDV3SDT', `Core10SDV2SDTV3SDTQ', and `Core18SV2SDV3SDT' 
	GAS models are adopted for both DC and RECP with small core Hamiltonians for the MRCI calculations. In the large-core RECP calculation, the 
	outer lyingelectrons are limited to five, which is undertaken by the `V5SDTQ' GAS model, where `V5' represents for the $ns^2np^3$ valence 
	electrons.

	\subsection{Evaluation of energies and properties}
	
	We estimate the energies with CBS limit ($E_{\rm{CBS}}$), then the net energy $E_X$ is given by 
	\begin{equation}
		E_X=E_{CBS}+AX^{-3},
		\label{eq:dyallcbs}
	\end{equation}	
	where $X$ is the cordinal numbers for the basis sets and $A$ is a fitting parameter as defined in Ref. \cite{Helgaker-JCP-1997}.
	
	The properties of our interest in this study are the lifetimes, Land{\' e} $g_J$ factors, and hyperfine structure constants of the atomic states. Since we are only studying fine-structure splitting partners of the ground state configurations of the undertaken atomic systems, the decay channels of the excited states will mainly be carried out through the M1 and electric quadrupole (E2) forbidden transitions. Thus, determination of lifetimes of the considered atomic states require transition probabilities due to M1 and E2 channels for which it is necessary to evaluate the corresponding transition amplitudes. 
	
	The transition matrix element due to M1 operator is calculated as  
	\begin{equation}
		\label{eq:M1}	
		\langle M1 \rangle=\frac{1}{C_1}\bigg\langle \psi_{J \Omega_J} \bigg|\sum_i^{N} \big(\bm{\vec \alpha}_i \times {\vec {\bf r}}_i \big)_z \bigg | \psi_{J' \Omega_{J'}} \bigg \rangle,
	\end{equation}
	where $J, J'$ denote the electronic angular momentum, and $\Omega_{J,J'}$ are their the projected values. 
	
	The E2 transition matrix element is calculated as
	\begin{eqnarray}
		\label{eq:E2}
		\langle E2 \rangle&=&\frac{ {(-1)^{(J-\Omega_J)}}}{C_2}\nonumber \\
		&&\bigg \langle \psi_{J \Omega_J} \bigg| \sum_i^{N}-\frac{3}{2}\big( {\vec {\bf r}}_{i\alpha} {\vec {\bf r}}_{i\beta}-\frac{1}{3}
		\delta_{\alpha\beta}r^2_i \big )_{zz} \bigg| \psi_{J' \Omega_{J'}} \bigg \rangle, \nonumber \\
	\end{eqnarray}
	where $\alpha,\beta=x,y,z$, $C_1$ and $C_2$ are the $3j$-symbol $
	\Bigg( \begin{array}{c c c} J&1&J'\\-\Omega_J&0&\Omega_{J'} \end{array} \Bigg) $ and $ 
	\Bigg( \begin{array}{c c c} J&2&J'\\-\Omega_J&0&\Omega_{J'} \end{array} \Bigg) $, respectively. 
	
	Then, the lifetime $\tau$ of the state within $np^3$ configuration (in second) is obtained as
	\begin{eqnarray}
		\tau=\frac{1}{\sum_{i,O} A_{fi}^O},
		\label{eq:tau}
	\end{eqnarray}
	where $\lambda_{fi}$ (in {\AA}) is the transition wavelength between the upper state $f$ and the down state $i$, and $A_{fi}^O$ is 
	\begin{eqnarray}
		A_{fi}^{M1} = \frac{2.69735\times10^{13}}{(2J_f+1)\lambda_{fi}^3}S_{fi}^{M1}, 
	\end{eqnarray}	
	and
	\begin{eqnarray}	 
		A_{fi}^{E2} = \frac{1.11995\times10^{18}}{(2J_f+1)\lambda_{fi}^5}S_{fi}^{E2},
	\end{eqnarray} 
	for the line strength $S_{fi}^O=$ (in a.u). 
	
	To determine the Dirac value of the Land\'{e} $g_J$ factor, an expectation value of a magnetic dipole operator is calculated by \cite{Yu-PRA-2020}
	\begin{equation} \label{gd}
		g_J^D =\frac{1}{\Omega_J}\bigg \langle \psi_{J \Omega_J} \bigg|\sum_i \big(\bm{\vec \alpha}_i \times {\vec {\bf r}}_i \big)_z \bigg| \psi_{J \Omega_{J}} \bigg \rangle.
	\end{equation}
	
	It is also known in the case of a free electron that QED corrections contribute significantly to the $g_J$ factor. To include the dominant 
	contributions from the QED effects, we estimate these corrections approximately as 
	\begin{equation} \label{gqed}
		\Delta g_J^Q=\frac{g_e-2}{2\Omega_J} \bigg \langle\psi_{J \Omega_J} \bigg | \sum_i \big(\bm \beta \bm \Sigma_z \big)_i \bigg | \psi_{J \Omega_J} \bigg \rangle,
	\end{equation}
	where $\bm \Sigma$ is the $4\times4$ spin matrix, and $g_e=2.002 319 3$ is the free-electron Land\'{e} $g$-factor. Therefore, the net value to the bound electron Land\'{e} $g_J$ factor is given by
	\begin{equation}
		g_J^{Total}=g_J^D+\Delta g_J^Q.
		\label{eq:gJ}
	\end{equation}
	
	Precise evaluation of hyperfine structure constants will also test the potential of method to determine atomic wave functions within the nuclear 
	region accurately. The $A$ value is estimated as the expectation of the M1 hyperfine operaor given by \cite{Fleig-JMS-2014}
	\begin{equation} \label{HFSA}
		A =\frac{\mu_N}{I\Omega_j}\bigg \langle\psi_{J \Omega_J} \bigg | \sum_i^N \bigg (\frac{\bm{\vec \alpha}_i \times {\vec {\bf r}}_i }
		{r^3_i}\bigg )_z \bigg |\psi_{J \Omega_J} \bigg \rangle,
	\end{equation}
	where $\mu_N$ and $I$ are the nuclear magnetic moment and the nuclear spin quantum number respectively. Similarly, the $B$ constant is calculated
	as the expectation value of the electric-field gradient tensor operator given by \cite{Visscher-JCP-1998}
	\begin{equation} \label{HyfA-2}
		B={Q}\bigg \langle \psi_{JJ} \bigg| \sum_i^N -\bigg (\frac{ 3 {\vec {\bf r}}_{i\alpha} {\vec {\bf r}}_{i\beta}-\delta_{\alpha\beta}
			r^2_i   }{r^5_i}\bigg )_{zz}\bigg|\psi_{JJ} \bigg \rangle,
	\end{equation}
	where $Q$ is the nuclear electric quadrupole moment and $\psi_{JJ}$ is known as the stretch state with $\Omega_J=J$.

	\begin{table*}[btp] 
		\caption{Energies (in cm$^{-1}$) of the excited states in the $np^3$ configuration ($n=3-6$) obtained using the DC Hamiltonian in the MRCI method with
			dyall.aaeXz basis sets adopted, $X$ = 2, 3, and 4. Differences among `Final' and `NIST' values are quoted as `Diff.'. \label{tab:dyall}}
		{\setlength{\tabcolsep}{0.7pt} 
			\begin{tabular}{lrrrr rrrrr rrrrr}\hline\hline 
				\addlinespace[0.2cm]
				\multirow{2}{*}{n=3}	&	\multicolumn{4}{c}{P}							&&	\multicolumn{4}{c}{S$^+$}							&&	\multicolumn{4}{c}{Cl$^{2+}$}							\\ \cline{2-5}\cline{7-10}\cline{12-15} \addlinespace[0.1cm]
				&	$^2D_{3/2}$	&	$^2D_{5/2}$	&	$^2P_{1/2}$	&	$^2P_{3/2}$	&&	$^2D_{3/2}$	&	$^2D_{5/2}$	&	$^2P_{1/2}$	&	$^2P_{3/2}$	&&	$^2D_{3/2}$	&	$^2D_{5/2}$	&	$^2P_{1/2}$	&	$^2P_{3/2}$	\\ \hline \addlinespace[0.1cm]
				$X$=2	&	13478 	&	13499 	&	20663 	&	20690 	&&	17307 	&	17348 	&	26803 	&	26854 	&&	20838 	&	20918 	&	32508 	&	32606 	\\
				$X$=3	&	11934 	&	11953 	&	19631 	&	19659 	&&	15542 	&	15581 	&	25582 	&	25632 	&&	18825 	&	18902 	&	30965 	&	31064 	\\
				$X$=4	&	11643 	&	11662 	&	19305 	&	19332 	&&	15189 	&	15228 	&	25179 	&	25229 	&&	18439 	&	18517 	&	30535 	&	30635 	\\
				$E_{CBS}$	&	11345 	&	11364 	&	19143 	&	19171 	&&	14854 	&	14894 	&	24992 	&	25042 	&&	18052 	&	18129 	&	30277 	&	30375 	\\
				$\Delta_{cor}$	&	$-$104 	&	$-$104 	&	$-$21 	&	$-$18 	&&	$-$105 	&	$-$105 	&	$-$27 	&	$-$25 	&&	$-$106 	&	$-$107 	&	$-$49 	&	$-$47 	\\
				$\Delta_{virt}$	&	18 	&	18 	&	91 	&	91 	&&	7 	&	7 	&	88 	&	88 	&&	$-$5 	&	$-$5 	&	80 	&	80 	\\
				$\Delta_{basis}$	&	$-$298 	&	$-$297 	&	$-$162 	&	$-$161 	&&	$-$335 	&	$-$334 	&	$-$187 	&	$-$187 	&&	$-$387 	&	$-$388 	&	$-$258 	&	$-$260 	\\
				Final	&	11258 	&	11278 	&	19212 	&	19244 	&&	14756 	&	14795 	&	25053 	&	25105 	&&	17941 	&	18017 	&	30307 	&	30408 	\\
				Uncert.	&	316 	&	315 	&	187 	&	186 	&&	351 	&	350 	&	208 	&	209 	&&	402 	&	403 	&	275 	&	276 	\\
				NIST\cite{NIST}	&	11361.02	&	11376.63	&	18722.71	&	18748.01	&&	14852.94	&	14884.73	&	24524.83	&	24571.54	&&	18052.46	&	18118.43	&	29813.92	&	29906.45	\\
				Diff.(\%)	&	$-$0.90 	&	$-$0.87 	&	2.62 	&	2.65 	&&	$-$0.65 	&	$-$0.60 	&	2.15 	&	2.17 	&&	$-$0.62 	&	$-$0.56 	&	1.65 	&	1.68 	\\ \hline \addlinespace[0.2cm]
				\multirow{2}{*}{n=4}	&	\multicolumn{4}{c}{As}							&&	\multicolumn{4}{c}{Se$^+$}							&&	\multicolumn{4}{c}{Br$^{2+}$}							\\ \cline{2-5}\cline{7-10}\cline{12-15} \addlinespace[0.1cm]
				&	$^2D_{3/2}$	&	$^2D_{5/2}$	&	$^2P_{1/2}$	&	$^2P_{3/2}$	&&	$^2D_{3/2}$	&	$^2D_{5/2}$	&	$^2P_{1/2}$	&	$^2P_{3/2}$	&&	$^2D_{3/2}$	&	$^2D_{5/2}$	&	$^2P_{1/2}$	&	$^2P_{3/2}$	\\ \hline \addlinespace[0.1cm]
				$X$=2	&	12678 	&	12997 	&	20306 	&	20755 	&&	15551 	&	16167 	&	25483 	&	26326 	&&	17943 	&	19006 	&	30059 	&	31510 	\\
				$X$=3	&	11325 	&	11627 	&	19389 	&	19824 	&&	14079 	&	14663 	&	24499 	&	25311 	&&	16298 	&	17311 	&	28850 	&	30257 	\\
				$X$=4	&	11032 	&	11337 	&	19087 	&	19530 	&&	13720 	&	14310 	&	24104 	&	24930 	&&	15904 	&	16926 	&	28413 	&	29840 	\\
				$E_{CBS}$	&	10781 	&	11081 	&	18945 	&	19382 	&&	13458 	&	14038 	&	23973 	&	24785 	&&	15609 	&	16616 	&	28240 	&	29649 	\\
				$\Delta_{cor}$	&	$-$87 	&	$-$84 	&	44 	&	62 	&&	$-$108 	&	$-$106 	&	20 	&	39 	&&	$-$138 	&	$-$134 	&	$-$36 	&	$-$12 	\\
				$\Delta_{virt}$	&	$-$14 	&	$-$12 	&	46 	&	53 	&&	$-$40 	&	$-$35 	&	0 	&	10 	&&	$-$66 	&	$-$58 	&	$-$39 	&	$-$23 	\\
				$\Delta_{basis}$	&	$-$251 	&	$-$256 	&	$-$142 	&	$-$148 	&&	$-$261 	&	$-$272 	&	$-$131 	&	$-$145 	&&	$-$295 	&	$-$309 	&	$-$173 	&	$-$191 	\\
				Final	&	10680 	&	10985 	&	19036 	&	19496 	&&	13310 	&	13896 	&	23993 	&	24835 	&&	15405 	&	16425 	&	28164 	&	29614 	\\
				Uncert.	&	266 	&	270 	&	156 	&	169 	&&	286 	&	294 	&	132 	&	150 	&&	332 	&	342 	&	181 	&	193 	\\
				NIST\cite{NIST}	&	10592.5	&	10914.6	&	18186.1	&	18647.5	&&	13168.2	&	13784.4	&	23038.3	&	23894.8	&&	15042	&	16301	&	26915	&	28579	\\
				Diff.(\%)	&	0.82 	&	0.65 	&	4.67 	&	4.55 	&&	1.07 	&	0.81 	&	4.14 	&	3.93 	&&	2.41 	&	0.76 	&	4.64 	&	3.62 	\\ \hline \addlinespace[0.2cm]
				\multirow{2}{*}{n=5}	&	\multicolumn{4}{c}{Sb}							&&	\multicolumn{4}{c}{Te$^+$}							&&	\multicolumn{4}{c}{I$^{2+}$}							\\ \cline{2-5}\cline{7-10}\cline{12-15} \addlinespace[0.1cm]
				&	$^2D_{3/2}$	&	$^2D_{5/2}$	&	$^2P_{1/2}$	&	$^2P_{3/2}$	&&	$^2D_{3/2}$	&	$^2D_{5/2}$	&	$^2P_{1/2}$	&	$^2P_{3/2}$	&&	$^2D_{3/2}$	&	$^2D_{5/2}$	&	$^2P_{1/2}$	&	$^2P_{3/2}$	\\ \hline \addlinespace[0.1cm]
				$X$=2	&	10755 	&	12076 	&	18857 	&	20829 	&&	12718 	&	14935 	&	23374 	&	26722 	&&	14376 	&	17661 	&	27489 	&	32637 	\\
				$X$=3	&	9368 	&	10617 	&	17873 	&	19804 	&&	11228 	&	13302 	&	22269 	&	25516 	&&	12739 	&	15817 	&	26148 	&	31180 	\\
				$X$=4	&	9011 	&	10280 	&	17446 	&	19423 	&&	10769 	&	12883 	&	21689 	&	25038 	&&	12266 	&	15377 	&	25506 	&	30664 	\\
				$E_{CBS}$	&	8770 	&	10015 	&	17323 	&	19276 	&&	10530 	&	12598 	&	21581 	&	24875 	&&	11994 	&	15045 	&	25356 	&	30450 	\\
				$\Delta_{cor}$	&	$-$178 	&	$-$206 	&	19 	&	49 	&&	$-$206 	&	$-$239 	&	$-$25 	&	1 	&&	$-$233 	&	$-$268 	&	$-$90 	&	$-$58 	\\
				$\Delta_{virt}$	&	28 	&	32 	&	66 	&	84 	&&	26 	&	34 	&	62 	&	82 	&&	24 	&	33 	&	59 	&	82 	\\
				$\Delta_{basis}$	&	$-$241 	&	$-$265 	&	$-$122 	&	$-$146 	&&	$-$238 	&	$-$285 	&	$-$108 	&	$-$163 	&&	$-$272 	&	$-$332 	&	$-$150 	&	$-$214 	\\
				Final	&	8620 	&	9841 	&	17408 	&	19410 	&&	10351 	&	12392 	&	21618 	&	24958 	&&	11786 	&	14809 	&	25325 	&	30474 	\\
				Uncert.	&	301 	&	337 	&	140 	&	176 	&&	316 	&	374 	&	127 	&	182 	&&	359 	&	428 	&	185 	&	236 	\\
				NIST\cite{NIST}	&	8512.125	&	9854.018	&	16395.359	&	18464.202	&&	10222.385	&	12421.854	&	20546.591	&	24032.095	&&	11710.6	&	14901	&	24298.8	&	29636.6	\\
				Diff.(\%)	&	1.26 	&	$-$0.13 	&	6.18 	&	5.12 	&&	1.26 	&	$-$0.24 	&	5.21 	&	3.85 	&&	0.64 	&	$-$0.61 	&	4.22 	&	2.82 	\\ \hline \addlinespace[0.2cm]
				\multirow{2}{*}{n=6}	&	\multicolumn{4}{c}{Bi}							&&	\multicolumn{4}{c}{Po$^+$}							&&	\multicolumn{4}{c}{At$^{2+}$}							\\ \cline{2-5}\cline{7-10}\cline{12-15} \addlinespace[0.1cm]
				&	$^2D_{3/2}$	&	$^2D_{5/2}$	&	$^2P_{1/2}$	&	$^2P_{3/2}$	&&	$^2D_{3/2}$	&	$^2D_{5/2}$	&	$^2P_{1/2}$	&	$^2P_{3/2}$	&&	$^2D_{3/2}$	&	$^2D_{5/2}$	&	$^2P_{1/2}$	&	$^2P_{3/2}$	\\ \hline \addlinespace[0.1cm]
				$X$=2	&	11857 	&	16446 	&	22966 	&	34059 	&&	16970 	&	22932 	&	31093 	&	47895 	&&	22971 	&	30069 	&	39562 	&	62911 	\\
				$X$=3	&	11401 	&	15611 	&	22444 	&	33564 	&&	16672 	&	22107 	&	30597 	&	47404 	&&	22750 	&	29164 	&	38891 	&	62227 	\\
				$X$=4	&	11416 	&	15536 	&	22299 	&	33626 	&&	16724 	&	21993 	&	30353 	&	47353 	&&	22879 	&	29080 	&	38648 	&	62212 	\\
				$E_{CBS}$	&	11299 	&	15352 	&	22211 	&	33487 	&&	16636 	&	21822 	&	30301 	&	47246 	&&	22786 	&	28878 	&	38549 	&	62045 	\\
				$\Delta_{cor}$	&	$-$107 	&	$-$191 	&	$-$113 	&	$-$110 	&&	$-$85 	&	$-$184 	&	$-$188 	&	$-$159 	&&	$-$57 	&	$-$163 	&	$-$209 	&	$-$149 	\\
				$\Delta_{virt}$	&	133 	&	172 	&	153 	&	323 	&&	124 	&	184 	&	124 	&	284 	&&	121 	&	170 	&	118 	&	284 	\\
				$\Delta_{basis}$	&	$-$117 	&	$-$184 	&	$-$88 	&	$-$139 	&&	$-$87 	&	$-$171 	&	$-$52 	&	$-$106 	&&	$-$94 	&	$-$201 	&	$-$99 	&	$-$167 	\\
				Final	&	11324 	&	15333 	&	22251 	&	33701 	&&	16675 	&	21822 	&	30237 	&	47371 	&&	22850 	&	28885 	&	38457 	&	62180 	\\
				Uncert.	&	207 	&	316 	&	210 	&	368 	&&	173 	&	311 	&	231 	&	342 	&&	163 	&	310 	&	260 	&	361 	\\
				NIST\cite{NIST}	&	11419.039	&	15437.501	&	21660.914	&	33164.805	&&		&		&		&		&&		&		&		&		\\
				Diff.(\%)	&	$-$0.84 	&	$-$0.68 	&	2.72 	&	1.62 	&&		&		&		&		&&		&		&		&		\\ \hline\hline
		\end{tabular}}
	\end{table*}
	
	\begin{table*}[btp]
		\caption{Energies (in cm$^{-1}$) of the excited states in the $np^3$ configuration ($n=3-4$) obtained using the DC Hamiltonian in the MRCI method with 
			aug-cc-pVXZ basis sets adopted, $X$ = 3, 4, and 5. Differences among the `Final' and `NIST' values are shown as `Diff.'. \label{AEperterson}}
		{\setlength{\tabcolsep}{2pt} 
			\begin{tabular}{lrrrr rrrrr rrrrr}\hline\hline \addlinespace[0.2cm]
				\multirow{2}{*}{n=3}	&	\multicolumn{4}{c}{P}							&&	\multicolumn{4}{c}{S$^+$}							&&	\multicolumn{4}{c}{Cl$^{2+}$}							\\ \cline{2-5}\cline{7-10}\cline{12-15} \addlinespace[0.1cm]
				&	$^2D_{3/2}$	&	$^2D_{5/2}$	&	$^2P_{1/2}$	&	$^2P_{3/2}$	&&	$^2D_{3/2}$	&	$^2D_{5/2}$	&	$^2P_{1/2}$	&	$^2P_{3/2}$	&&	$^2D_{3/2}$	&	$^2D_{5/2}$	&	$^2P_{1/2}$	&	$^2P_{3/2}$	\\ \hline \addlinespace[0.1cm]
				$X$=5	&	11584 	&	11603 	&	19244 	&	19272 	&&	15115 	&	15154 	&	25103 	&	25155 	&&	18359 	&	18436 	&	30459 	&	30560 	\\
				$X$=6	&	11549 	&	11568 	&	19238 	&	19265 	&&	15066 	&	15105 	&	25078 	&	25130 	&&	18274 	&	18352 	&	30390 	&	30492 	\\
				$X$=7	&	11544 	&	11563 	&	19184 	&	19212 	&&	15072 	&	15111 	&	25035 	&	25086 	&&	18320 	&	18399 	&	30390 	&	30490 	\\
				$E_{CBS}$	&	11524 	&	11543 	&	19178 	&	19205 	&&	15044 	&	15083 	&	25019 	&	25069 	&&	18273 	&	18353 	&	30350 	&	30451 	\\
				$\Delta_{cor}$	&	$-$51 	&	$-$51 	&	$-$128 	&	$-$128 	&&	$-$49 	&	$-$48 	&	$-$111 	&	$-$110 	&&	$-$67 	&	$-$67 	&	$-$137 	&	$-$138 	\\
				$\Delta_{basis}$	&	$-$20 	&	$-$20 	&	$-$6 	&	$-$7 	&&	$-$28 	&	$-$28 	&	$-$16 	&	$-$17 	&&	$-$47 	&	$-$46 	&	$-$40 	&	$-$39 	\\
				Final	&	11473 	&	11492 	&	19050 	&	19077 	&&	14995 	&	15035 	&	24908 	&	24959 	&&	18206 	&	18286 	&	30213 	&	30313 	\\
				Uncert.	&	55 	&	55 	&	128 	&	128 	&&	57 	&	55 	&	112 	&	111 	&&	82 	&	81 	&	143 	&	143 	\\
				NIST\cite{NIST}	&	11361.02	&	11376.63	&	18722.71	&	18748.01	&&	14852.94	&	14884.73	&	24524.83	&	24571.54	&&	18052.46	&	18118.43	&	29813.92	&	29906.45	\\
				Diff.(\%)	&	0.99 	&	1.01 	&	1.75 	&	1.75 	&&	0.96 	&	1.01 	&	1.56 	&	1.58 	&&	0.85 	&	0.92 	&	1.34 	&	1.36 	\\ \hline \addlinespace[0.2cm]
				\multirow{2}{*}{n=4}	&	\multicolumn{4}{c}{As}							&&	\multicolumn{4}{c}{Se$^+$}							&&	\multicolumn{4}{c}{Br$^{2+}$}							\\ \cline{2-5}\cline{7-10}\cline{12-15} \addlinespace[0.1cm]
				&	$^2D_{3/2}$	&	$^2D_{5/2}$	&	$^2P_{1/2}$	&	$^2P_{3/2}$	&&	$^2D_{3/2}$	&	$^2D_{5/2}$	&	$^2P_{1/2}$	&	$^2P_{3/2}$	&&	$^2D_{3/2}$	&	$^2D_{5/2}$	&	$^2P_{1/2}$	&	$^2P_{3/2}$	\\ \hline \addlinespace[0.1cm]
				$X$=3	&	11361 	&	11580 	&	19164 	&	19471 	&&	14226 	&	14648 	&	24278 	&	24855 	&&	16533 	&	17276 	&	28586 	&	29603 	\\
				$X$=4	&	11038 	&	11252 	&	18820 	&	19132 	&&	13786 	&	14214 	&	23800 	&	24391 	&&	16071 	&	16816 	&	28106 	&	29133 	\\
				$X$=5	&	10919 	&	11172 	&	18778 	&	19140 	&&	13604 	&	14089 	&	23695 	&	24365 	&&	15809 	&	16649 	&	27929 	&	29085 	\\
				$E_{CBS}$	&	10800 	&	11046 	&	18640 	&	18998 	&&	13443 	&	13923 	&	23508 	&	24175 	&&	15646 	&	16478 	&	27750 	&	28894 	\\
				$\Delta_{cor}$	&	$-$38 	&	$-$35 	&	$-$63 	&	$-$53 	&&	$-$31 	&	$-$27 	&	$-$41 	&	$-$31 	&&	$-$37 	&	$-$32 	&	$-$54 	&	$-$41 	\\
				$\Delta_{basis}$	&	$-$119 	&	$-$126 	&	$-$138 	&	$-$142 	&&	$-$161 	&	$-$166 	&	$-$187 	&	$-$190 	&&	$-$162 	&	$-$171 	&	$-$179 	&	$-$191 	\\
				Final	&	10762 	&	11011 	&	18577 	&	18945 	&&	13412 	&	13896 	&	23467 	&	24145 	&&	15610 	&	16446 	&	27697 	&	28852 	\\
				Uncert.	&	125 	&	131 	&	152 	&	152 	&&	164 	&	168 	&	191 	&	192 	&&	167 	&	174 	&	187 	&	196 	\\
				NIST\cite{NIST}	&	10592.5	&	10914.6	&	18186.1	&	18647.5	&&	13168.2	&	13784.4	&	23038.3	&	23894.8	&&	15042	&	16301	&	26915	&	28579	\\
				Diff.(\%)	&	1.60 	&	0.89 	&	2.15 	&	1.59 	&&	1.85 	&	0.81 	&	1.86 	&	1.05 	&&	3.77 	&	0.89 	&	2.90 	&	0.96 	\\ \hline\hline
		\end{tabular}}
	\end{table*}
	
	\begin{table*}[btp]
		\caption{Energies (in cm$^{-1}$) of the excited states in the $np^3$ configuration ($n=4-6$) obtained by using the small-core RECP potential in the MRCI method with aug-cc-pVXZ-PP basis sets adopted, $X$ = 3, 4, and 5. Differences in the `Final' and `NIST' values are given as `Diff.'. \label{np-sECP-cc}}
		{\setlength{\tabcolsep}{1pt}
			\begin{tabular}{lrrrr rrrrr rrrrr}\hline\hline \addlinespace[0.2cm]
				\multirow{2}{*}{n=4}	&	\multicolumn{4}{c}{As}							&&	\multicolumn{4}{c}{Se$^+$}							&&	\multicolumn{4}{c}{Br$^{2+}$}							\\ \cline{2-5}\cline{7-10}\cline{12-15}  \addlinespace[0.1cm]
				&	$^2D_{3/2}$	&	$^2D_{5/2}$	&	$^2P_{1/2}$	&	$^2P_{3/2}$	&&	$^2D_{3/2}$	&	$^2D_{5/2}$	&	$^2P_{1/2}$	&	$^2P_{3/2}$	&&	$^2D_{3/2}$	&	$^2D_{5/2}$	&	$^2P_{1/2}$	&	$^2P_{3/2}$	\\ \hline \addlinespace[0.1cm]
				$X$=3	&	11315 	&	11576 	&	19157 	&	19529 	&&	14120 	&	14641 	&	24258 	&	24978 	&&	16373 	&	17298 	&	28600 	&	29878 	\\
				$X$=4	&	10986 	&	11258 	&	18866 	&	19255 	&&	13691 	&	14226 	&	23854 	&	24594 	&&	15890 	&	16833 	&	28139 	&	29444 	\\
				$X$=5	&	10896 	&	11177 	&	18822 	&	19225 	&&	13549 	&	14098 	&	23744 	&	24506 	&&	15708 	&	16661 	&	27943 	&	29274 	\\
				$E_{CBS}$	&	10770 	&	11055 	&	18706 	&	19116 	&&	13386 	&	13939 	&	23589 	&	24355 	&&	15528 	&	16487 	&	27775 	&	29113 	\\
				$\Delta_{cor}$	&	$-$34	&	$-$31	&	$-$72	&	$-$68	&&	$-$33	&	$-$30	&	$-$54	&	$-$48	&&	$-$33	&	$-$27	&	$-$40	&	$-$30	\\
				$\Delta_{basis}$	&	$-$126 	&	$-$122 	&	$-$116 	&	$-$109 	&&	$-$163 	&	$-$159 	&	$-$155 	&	$-$151 	&&	$-$180 	&	$-$174 	&	$-$168 	&	$-$161 	\\
				Final	&	10736 	&	11024 	&	18634 	&	19048 	&&	13353 	&	13909 	&	23535 	&	24307 	&&	15495 	&	16460 	&	27735 	&	29083 	\\
				Uncert.	&	131 	&	125 	&	137 	&	128 	&&	166 	&	162 	&	164 	&	159 	&&	183 	&	176 	&	173 	&	164 	\\
				NIST\cite{NIST}	&	10592.5	&	10914.6	&	18186.1	&	18647.5	&&	13168.2	&	13784.4	&	23038.3	&	23894.8	&&	15042	&	16301	&	26915	&	28579	\\
				Diff.(\%)	&	1.35 	&	1.00 	&	2.46 	&	2.15 	&&	1.40 	&	0.90 	&	2.16 	&	1.73 	&&	3.01 	&	0.97 	&	3.04 	&	1.76 	\\ \hline \addlinespace[0.1cm]
				\multirow{2}{*}{n=5}	&	\multicolumn{4}{c}{Sb }							&&	\multicolumn{4}{c}{Te$^+$}							&&	\multicolumn{4}{c}{I$^{2+}$}							\\ \cline{2-5}\cline{7-10}\cline{12-15} \addlinespace[0.1cm]
				&	$^2D_{3/2}$	&	$^2D_{5/2}$	&	$^2P_{1/2}$	&	$^2P_{3/2}$	&&	$^2D_{3/2}$	&	$^2D_{5/2}$	&	$^2P_{1/2}$	&	$^2P_{3/2}$	&&	$^2D_{3/2}$	&	$^2D_{5/2}$	&	$^2P_{1/2}$	&	$^2P_{3/2}$	\\ \hline \addlinespace[0.1cm]
				$X$=3	&	9419 	&	10473 	&	17382 	&	18979 	&&	11296 	&	13149 	&	21686 	&	24510 	&&	13076 	&	15901 	&	26117 	&	30499 	\\
				$X$=4	&	9047 	&	10190 	&	17179 	&	18913 	&&	10860 	&	12830 	&	21449 	&	24476 	&&	12362 	&	15326 	&	25250 	&	30021 	\\
				$X$=5	&	8931 	&	10149 	&	17170 	&	19036 	&&	10708 	&	12736 	&	21361 	&	24517 	&&	12201 	&	15236 	&	25145 	&	30094 	\\
				$E_{CBS}$	&	8790 	&	10037 	&	17086 	&	18993 	&&	10544 	&	12613 	&	21271 	&	24498 	&&	11924 	&	15008 	&	24796 	&	29888 	\\
				$\Delta_{cor}$	&	$-$39 	&	$-$31	&	$-$87	&	$-$73	&&	$-$39	&	$-$28	&	$-$72	&	$-$54	&&	$-$32	&	$-$21	&	$-$51	&	$-$33	\\
				$\Delta_{basis}$	&	$-$141 	&	$-$112 	&	$-$84 	&	$-$42 	&&	$-$164 	&	$-$123 	&	$-$89 	&	$-$19 	&&	$-$277 	&	$-$228 	&	$-$349 	&	$-$206 	\\
				Final	&	8751 	&	10006 	&	16999 	&	18920 	&&	10505 	&	12585 	&	21199 	&	24444 	&&	11892 	&	14987 	&	24745 	&	29855 	\\
				Uncert.	&	146 	&	116 	&	121 	&	84 	&&	169 	&	126 	&	115 	&	57 	&&	279 	&	229 	&	352 	&	209 	\\
				NIST\cite{NIST}	&	8512.125	&	9854.018	&	16395.359	&	18464.202	&&	10222.385	&	12421.854	&	20546.591	&	24032.095	&&	11710.6	&	14901	&	24298.8	&	29636.6	\\
				Diff.(\%)	&	2.81 	&	1.54 	&	3.68 	&	2.47 	&&	2.76 	&	1.32 	&	3.18 	&	1.71 	&&	1.55 	&	0.57 	&	1.84 	&	0.74 	\\ \hline \addlinespace[0.1cm]
				\multirow{2}{*}{n=6}	&	\multicolumn{4}{c}{Bi}							&&	\multicolumn{4}{c}{Po$^+$}							&&	\multicolumn{4}{c}{At$^{2+}$}							\\ \cline{2-5}\cline{7-10}\cline{12-15} \addlinespace[0.1cm]
				&	$^2D_{3/2}$	&	$^2D_{5/2}$	&	$^2P_{1/2}$	&	$^2P_{3/2}$	&&	$^2D_{3/2}$	&	$^2D_{5/2}$	&	$^2P_{1/2}$	&	$^2P_{3/2}$	&&	$^2D_{3/2}$	&	$^2D_{5/2}$	&	$^2P_{1/2}$	&	$^2P_{3/2}$	\\ \hline \addlinespace[0.1cm]
				$X$=3	&	9884 	&	13709 	&	20322 	&	29034 	&&	14254 	&	19468 	&	27584 	&	41388 	&&	20072 	&	27288 	&	35969 	&	55652 	\\
				$X$=4	&	10824 	&	14834 	&	21562 	&	31938 	&&	15589 	&	20794 	&	29138 	&	44641 	&&	21256 	&	27475 	&	37043 	&	58713 	\\
				$X$=5	&	11055 	&	15088 	&	21810 	&	32585 	&&	16087 	&	21280 	&	29583 	&	45755 	&&	22002 	&	28152 	&	37698 	&	60250 	\\
				$E_{CBS}$	&	11419 	&	15526 	&	22296 	&	33718 	&&	16584 	&	21774 	&	30178 	&	46976 	&&	22404 	&	28148 	&	38066 	&	61339 	\\
				$\Delta_{cor}$	&	$-$3 	&	5	&	$-$60	&	$-$6	&&	11	&	10	&	$-$50	&	$-$12	&&	34	&	31	&	$-$33	&	29	\\
				$\Delta_{basis}$	&	364 	&	438 	&	487 	&	1133 	&&	497 	&	494 	&	595 	&	1221 	&&	402 	&	$-$5 	&	367 	&	1089 	\\
				Final	&	11416 	&	15531 	&	22236 	&	33712 	&&	16595 	&	21784 	&	30128 	&	46964 	&&	22438 	&	28179 	&	38033 	&	61368 	\\
				Uncert.	&	364 	&	438 	&	491 	&	1133 	&&	497 	&	494 	&	597 	&	1221 	&&	403 	&	31 	&	369 	&	1089 	\\
				NIST\cite{NIST}	&	11419.039	&	15437.501	&	21660.914	&	33164.805	&&		&		&		&		&&		&		&		&		\\
				Diff.(\%)	&	$-$0.02 	&	0.60 	&	2.66 	&	1.65 	&&		&		&		&		&&		&		&		&		\\ \hline \hline
		\end{tabular}}
	\end{table*}
	
	\begin{table*}[btp]
		\caption{Energies (in cm$^{-1}$) of the excited states in the $np^3$ configuration ($n=4-6$) obtained using the large-core RECP potential in the MRCI method 
			with aug-cc-pVXZ-PP basis sets adopted, $X$ = 3, 4, and 5. Differences between `Final' and `NIST' results are given as `Diff.'. \label{np-lECP-cc}}
		{\setlength{\tabcolsep}{1pt}
			\begin{tabular}{lrrrr rrrrr rrrrr}\hline\hline \addlinespace[0.2cm]
				\multirow{2}{*}{n=4}	&	\multicolumn{4}{c}{As}							&&	\multicolumn{4}{c}{Se$^+$}							&&	\multicolumn{4}{c}{Br$^{2+}$}							\\ \cline{2-5}\cline{7-10}\cline{12-15}  \addlinespace[0.1cm]
				&	$^2D_{3/2}$	&	$^2D_{5/2}$	&	$^2P_{1/2}$	&	$^2P_{3/2}$	&&	$^2D_{3/2}$	&	$^2D_{5/2}$	&	$^2P_{1/2}$	&	$^2P_{3/2}$	&&	$^2D_{3/2}$	&	$^2D_{5/2}$	&	$^2P_{1/2}$	&	$^2P_{3/2}$	\\ \hline  \addlinespace[0.1cm]
				$X$=3	&	10984 	&	11183 	&	18637 	&	18923 	&&	14032 	&	14537 	&	23996 	&	24674 	&&	16771 	&	16981 	&	28255 	&	28547 	\\
				$X$=4	&	10616 	&	10818 	&	18113 	&	18410 	&&	13757 	&	14245 	&	23708 	&	24371 	&&	16489 	&	16793 	&	28121 	&	28549 	\\
				$X$=5	&	10795 	&	10967 	&	18348 	&	18599 	&&	13777 	&	14235 	&	23732 	&	24355 	&&	16179 	&	16419 	&	27611 	&	27946 	\\
				$E_{CBS}$	&	10620 	&	10796 	&	18100 	&	18361 	&&	13704 	&	14167 	&	23655 	&	24285 	&&	16098 	&	16386 	&	27616 	&	28020 	\\
				$\Delta_{basis}$	&	$-$175 	&	$-$171 	&	$-$248 	&	$-$238 	&&	$-$73 	&	$-$68 	&	$-$77 	&	$-$70 	&&	$-$81 	&	$-$33 	&	6 	&	75 	\\
				NIST\cite{NIST}	&	10592.5	&	10914.6	&	18186.1	&	18647.5	&&	13168.2	&	13784.4	&	23038.3	&	23894.8	&&	15042	&	16301	&	26915	&	28579	\\
				Diff.(\%)	&	0.26 	&	$-$1.09 	&	$-$0.47 	&	$-$1.54 	&&	4.07 	&	2.78 	&	2.68 	&	1.63 	&&	7.02 	&	0.52 	&	2.61 	&	$-$1.95 	\\ \hline  \addlinespace[0.2cm]
				\multirow{2}{*}{n=5}	&	\multicolumn{4}{c}{Sb }							&&	\multicolumn{4}{c}{Te$^+$}							&&	\multicolumn{4}{c}{I$^{2+}$}							\\ \cline{2-5}\cline{7-10}\cline{12-15}  \addlinespace[0.1cm]
				&	$^2D_{3/2}$	&	$^2D_{5/2}$	&	$^2P_{1/2}$	&	$^2P_{3/2}$	&&	$^2D_{3/2}$	&	$^2D_{5/2}$	&	$^2P_{1/2}$	&	$^2P_{3/2}$	&&	$^2D_{3/2}$	&	$^2D_{5/2}$	&	$^2P_{1/2}$	&	$^2P_{3/2}$	\\ \hline  \addlinespace[0.1cm]
				$X$=3	&	8818 	&	10084 	&	16618 	&	18521 	&&	10166 	&	12650 	&	20566 	&	24655 	&&	12811 	&	15411 	&	25541 	&	29522 	\\
				$X$=4	&	9044 	&	10108 	&	16821 	&	18396 	&&	11133 	&	12877 	&	21521 	&	24129 	&&	12646 	&	15263 	&	25289 	&	29353 	\\
				$X$=5	&	8982 	&	10032 	&	16707 	&	18262 	&&	11000 	&	12746 	&	21286 	&	23903 	&&	12523 	&	15325 	&	25297 	&	29722 	\\
				$E_{CBS}$	&	9084 	&	10052 	&	16805 	&	18225 	&&	11419 	&	12857 	&	21713 	&	23712 	&&	12468 	&	15256 	&	25191 	&	29607 	\\
				$\Delta_{basis}$	&	102 	&	20 	&	98 	&	$-$37 	&&	419 	&	110 	&	427 	&	$-$191 	&&	$-$54 	&	$-$69 	&	$-$106 	&	$-$115 	\\
				NIST\cite{NIST}	&	8512.125	&	9854.018	&	16395.359	&	18464.202	&&	10222.385	&	12421.854	&	20546.591	&	24032.095	&&	11710.6	&	14901	&	24298.8	&	29636.6	\\
				Diff.(\%)	&	6.72 	&	2.01 	&	2.50 	&	$-$1.29 	&&	11.71 	&	3.50 	&	5.68 	&	$-$1.33 	&&	6.47 	&	2.38 	&	3.67 	&	$-$0.10 	\\ \hline  \addlinespace[0.2cm]
				\multirow{2}{*}{n=6}	&	\multicolumn{4}{c}{Bi}							&&	\multicolumn{4}{c}{Po$^+$}							&&	\multicolumn{4}{c}{At$^{2+}$}							\\ \cline{2-5}\cline{7-10}\cline{12-15}  \addlinespace[0.1cm]
				&	$^2D_{3/2}$	&	$^2D_{5/2}$	&	$^2P_{1/2}$	&	$^2P_{3/2}$	&&	$^2D_{3/2}$	&	$^2D_{5/2}$	&	$^2P_{1/2}$	&	$^2P_{3/2}$	&&	$^2D_{3/2}$	&	$^2D_{5/2}$	&	$^2P_{1/2}$	&	$^2P_{3/2}$	\\ \hline  \addlinespace[0.1cm]
				$X$=3	&	10727 	&	14681 	&	20862 	&	31235 	&&	16493 	&	21706 	&	29411 	&	46393 	&&	22476 	&	29759 	&	38145 	&	60854 	\\
				$X$=4	&	10970 	&	14974 	&	21213 	&	31926 	&&	15583 	&	20827 	&	29133 	&	44518 	&&	21370 	&	27648 	&	37247 	&	58806 	\\
				$X$=5	&	10956 	&	14931 	&	21138 	&	31889 	&&	15796 	&	21036 	&	29262 	&	44994 	&&	21654 	&	27889 	&	37473 	&	59435 	\\
				$E_{CBS}$	&	11059 	&	15060 	&	21296 	&	32184 	&&	15390 	&	20644 	&	29129 	&	44154 	&&	21160 	&	26980 	&	37071 	&	58505 	\\
				$\Delta_{basis}$	&	103 	&	129 	&	157 	&	294 	&&	$-$406 	&	$-$392 	&	$-$133 	&	$-$840 	&&	$-$495 	&	$-$909 	&	$-$402 	&	$-$930 	\\
				NIST\cite{NIST}	&	11419.039	&	15437.501	&	21660.914	&	33164.805	&&		&		&		&		&&		&		&		&		\\
				Diff.(\%)	&	$-$3.15 	&	$-$2.44 	&	$-$1.69 	&	$-$2.96 	&&		&		&		&		&&		&		&		&		\\ \hline \hline
		\end{tabular}}
	\end{table*}
	
	\begin{table*}[h]
		\caption{Energies (in cm$^{-1}$) and $g_J$ factors of the ground and the low-lying excited states of Mc, Lv$^+$ and Ts$^{2+}$ obtained by using the DC and RECP Hamiltonians in the MRCI method and comparison with the CI+MBPT results \cite{Dzuba-HI-2016}. The dyall.aae4z (`aae4z') and the doubly-density-intensified dyall.aae4z
			basis set (`aae4z$^{+DD}$') are adopted for the DC Hamiltonian calculation, while the QZVP and dyall.aae4z basis sets are adopted for the RECP 
			calculations. The numbers in brackets are the uncertainties. Non-relativistic results for $g_J$ factors are also given as $g_J^{LS}$. \label{7p-EE}}
		{\setlength{\tabcolsep}{4pt}
			\begin{tabular}{lllrrrrrrrrrrrrr}\hline\hline \addlinespace[0.2cm]
				System	&	\multicolumn{2}{c}{State}			&	$g_J^{LS}$	&	\multicolumn{6}{c}{Energy (DC)}											&	&	\multicolumn{2}{c}{Energy (RECP)}			&	&	\multicolumn{2}{c}{CI+MBPT \cite{Dzuba-HI-2016}}			\\ \cline{2-3} \cline{5-10} \cline{12-13}\cline{15-16} \addlinespace[0.1cm]
				&		&		&		&	aae4z	&	aae4z$^{+DD}$	&	$\Delta_{DD}$	&	$\Delta_{cor}$	&	Final	&	$g_J$	&	&	QZVP	&	aae4z	&	&	EE	&	$g_J$	\\ \hline \addlinespace[0.1cm]
				Mc	&	$7p^3$	&	$^2P^o_{3/2}$	&	1.3333 	&	0 	&	0	&	0 	&	0 	&		&	1.3923 	&	&	0	&	0	&	&	0 	&	1.4039 	\\
				&	$7p^28s$	&	$^2S_{1/2}$	&	2.0000 	&	18533 	&	18525 	&	$-$8 	&	62 	&	18587(63)	&	2.0302 	&	&	47749 	&	18800 	&	&	18377 	&	2.0427 	\\
				&	$7p^28p$	&	$^2P^o_{1/2}$	&	0.6667 	&	26964 	&	26946 	&	$-$17 	&	10 	&	26957(10)	&	0.6668 	&	&	29717 	&	27254 	&	&	28382 	&	0.6668 	\\
				&	$7p^28p$	&	$^2P^o_{3/2}$	&	1.3333 	&	30534 	&	30517 	&	$-$16 	&	79 	&	30597(81)	&	1.3395 	&	&	36451 	&	30767 	&	&	31716 	&	1.3449 	\\
				&	$7p^27d$	&	$^2D_{3/2}$	&	0.8000 	&	32515 	&	32513 	&	$-$2 	&	41 	&	32554(41)	&	0.7969 	&	&	79103 	&	32766 	&	&	33092 	&	0.7987 	\\
				&	$7p^27d$	&	$^2D_{5/2}$	&	1.2000 	&	33240 	&	33237 	&	$-$3 	&	61 	&	33299(61)	&	1.2008 	&	&	83402 	&	33492 	&	&	33273 	&	1.2007 	\\
				&	$7p^3$	&	$^2P^o_{3/2}$	&	1.3333 	&	37289 	&	36785 	&	$-$504 	&	$-$171 	&	36615(532)	&	1.4380 	&	&	40041 	&	36667 	&	&	35484 	&	1.4295 	\\
				&	$7p^3$	&	$^2D_{5/2}$	&	1.2000 	&	41282 	&	40795 	&	$-$487 	&	568 	&	41364(748)	&	1.2013 	&	&	41366 	&	40637 	&	&		&		\\
				&	$7p^3$	&	$^2P^o_{1/2}$	&	0.6667 	&	47123 	&	46938 	&	$-$185 	&	$-$58 	&	46879(194)	&	0.6663 	&	&	47262 	&	46483 	&	&		&		\\ \hline \addlinespace[0.2cm]
				Lv$^+$	&	$7p^3$	&	$^2P^o_{3/2}$	&	1.3333 	&	0 	&	0 	&	0 	&	0 	&	0 	&	1.3834 	&	&	0 	&	0 	&	&		&		\\
				&	$7p^28s$	&	$^2S_{1/2}$	&	2.0000 	&	40564 	&	40398 	&	$-$166 	&	235 	&	40633(288)	&	2.0308 	&	&	57990 	&	40275 	&	&		&		\\
				&	$7p^3$	&	$^2P^o_{3/2}$	&	1.3333 	&	53505 	&	53517 	&	13 	&	$-$221 	&	53296(221)	&	1.4525 	&	&	54303 	&	54136 	&	&		&		\\
				&	$7p^28p$	&	$^2P^o_{1/2}$	&	0.6667 	&	56295 	&	56290 	&	$-$5 	&	207 	&	56497(207)	&	0.6669 	&	&	57537 	&	56169 	&	&		&		\\
				&	$7p^3$	&	$^2D^o_{5/2}$	&	1.2000 	&	58297 	&	58309 	&	12 	&	$-$108 	&	58201(108)	&	1.2006 	&	&	59454 	&	58359 	&	&		&		\\
				&	$7p^27d$	&	$^2D_{3/2}$	&	0.8000 	&	58530 	&	58527 	&	$-$3 	&	273 	&	58800(273)	&	0.7941 	&	&	84358 	&	58886 	&	&		&		\\
				&	$7p^27d$	&	$^2D_{5/2}$	&	1.2000 	&	61260 	&	61252 	&	$-$8 	&	331 	&	61582(331)	&	1.2020 	&	&	93214 	&	60941 	&	&		&		\\
				&	$7p^3$	&	$^2P^o_{1/2}$	&	0.6667 	&	65821 	&	65833 	&	12 	&	$-$147 	&	65686(148)	&	0.6684 	&	&	66954 	&	65833 	&	&		&		\\
				&	$7p^28p$	&	$^2P^o_{3/2}$	&	1.3333 	&	65869 	&	65844 	&	$-$25 	&	146 	&	65990(148)	&	1.3329 	&	&	72513 	&	66334 	&	&		&		\\ \hline \addlinespace[0.2cm]
				Ts$^{2+}$	&	$7p^3$	&	$^2P^o_{3/2}$	&	1.3333 	&	0 	&	0 	&	0 	&	0 	&	0 	&	1.3751 	&	&	0 	&	0.0000 	&	&		&		\\
				&	$7p^28s$	&	$^2S_{1/2}$	&	2.0000 	&	62950 	&	62857 	&	$-$93 	&	337 	&	63194(349)	&	2.0283 	&	&	74690 	&	63537 	&	&		&		\\
				&	$7p^3$	&	$^2P^o_{3/2}$	&	1.3333 	&	71165 	&	71194 	&	29 	&	$-$247 	&	70947(249)	&	1.4554 	&	&	70358 	&	70007 	&	&		&		\\
				&	$7p^3$	&	$^2D^o_{5/2}$	&	1.2000 	&	76412 	&	76440 	&	28 	&	$-$161 	&	76278(164)	&	1.1998 	&	&	75925 	&	75230 	&	&		&		\\
				&	$7p^27d$	&	$^2D_{3/2}$	&	0.8000 	&	80457 	&	80473 	&	16 	&	409 	&	80882(410)	&	0.7925 	&	&	94513 	&	80589 	&	&		&		\\
				&	$7p^3$	&	$^2P^o_{1/2}$	&	0.6667 	&	84436 	&	84455 	&	20 	&	$-$45 	&	84410(49)	&	0.6666 	&	&	84165 	&	83510 	&	&		&		\\
				&	$7p^27d$	&	$^2D_{5/2}$	&	1.2000 	&	85897 	&	85920 	&	22 	&	475 	&	86395(476)	&	1.2002 	&	&	104306 	&	85859 	&	&		&		\\
				&	$7p^28p$	&	$^2P^o_{1/2}$	&	0.6667 	&	86344 	&	86359 	&	15 	&	150 	&	86510(151)	&	0.6668 	&	&	87552 	&	86507 	&	&		&		\\
				&	$7p^28p$	&	$^2P^o_{3/2}$	&	1.3333 	&	102032 	&	102044 	&	12 	&	211 	&	102255(211)	&	1.3346 	&	&	107832 	&	102369 	&	&		&		\\
				&	$7p^28p$	&	$^2F^o_{5/2}$	&	0.8571 	&	112631 	&	112649 	&	18 	&	237 	&	112886(2380	&	0.8510 	&	&		&	112563 	&	&		&		\\ \hline \hline
		\end{tabular}}
	\end{table*}
	
	\begin{table*}[btp]
		\caption{The $g_J$ factor for the $np^3$ $^4S_{3/2}$, $^2D_{3/2,5/2}$, and $^2P_{3/2,1/2}$ states ($n=3-4$) . \label{gj_1}}
		{\setlength{\tabcolsep}{16pt}
			\begin{tabular}{ll rrrrr}\hline\hline \addlinespace[0.2cm]
				System	&		&	\multicolumn{1}{c}{$^4S_{3/2}$}	&	\multicolumn{1}{c}{$^2D_{3/2}$}	&	\multicolumn{1}{c}{$^2D_{5/2}$}	&	\multicolumn{1}{c}{$^2P_{1/2}$}	&	\multicolumn{1}{c}{$^2P_{3/2}$}	\\ \hline \addlinespace[0.1cm]
				P 	&	$g_j^D$	&	1.999767 	&	0.800184 	&	1.200500 	&	0.666120 	&	1.333132 	\\
				&	$\Delta g_j^Q$	&	0.009276 	&	$-$0.001853 	&	0.001861 	&	$-$0.003097 	&	0.003091 	\\
				&	$\Delta_{cor}$	&	$-$0.000001 	&	0.000298 	&	$-$0.000313 	&	$-$0.000519 	&	0.000536 	\\
				&	$\Delta_{virt}$	&	$-$0.000004 	&	$-$0.000079 	&	0.000069 	&	$-$0.000083 	&	0.000086 	\\
				&	$g_j^{Total}$	&	2.009038 	&	0.798550 	&	1.202117 	&	0.662421 	&	1.336845 	\\
				&	Uncert.	&	0.000004 	&	0.000308 	&	0.000321 	&	0.000526 	&	0.000543 	\\ [+2ex]
				S$^{+}$	&	$g_j^D$	&	1.999579 	&	0.801475 	&	1.199954 	&	0.666360 	&	1.332132 	\\
				&	$\Delta g_j^Q$	&	0.009274 	&	$-$0.001841 	&	0.001856 	&	$-$0.003095 	&	0.003082 	\\
				&	$\Delta_{cor}$	&	$-$0.000001 	&	$-$0.000082 	&	0.000056 	&	$-$0.000014 	&	0.000040 	\\
				&	$\Delta_{virt}$	&	$-$0.000005 	&	$-$0.000050 	&	0.000035 	&	0.000016 	&	$-$0.000008 	\\
				&	$g_j^{Total}$	&	2.008847 	&	0.799502 	&	1.201901 	&	0.663267 	&	1.335247 	\\
				&	Uncert.	&	0.000006 	&	0.000096 	&	0.000066 	&	0.000022 	&	0.000041 	\\ [+2ex]
				Cl$^{2+}$	&	$g_j^D$	&	1.999293 	&	0.802877 	&	1.199844 	&	0.666449 	&	1.330797 	\\
				&	$\Delta g_j^Q$	&	0.009272 	&	$-$0.001828 	&	0.001855 	&	$-$0.003094 	&	0.003070 	\\
				&	$\Delta_{cor}$	&	$-$0.000002 	&	$-$0.000050 	&	0.000021 	&	0.000264 	&	$-$0.000235 	\\
				&	$\Delta_{virt}$	&	$-$0.000007 	&	$-$0.000031 	&	0.000007 	&	0.000002 	&	0.000013 	\\
				&	$g_j^{Total}$	&	2.008556 	&	0.800970 	&	1.201727 	&	0.663621 	&	1.333645 	\\
				&	Uncert.	&	0.000008 	&	0.000058 	&	0.000022 	&	0.000264 	&	0.000235 	\\ [+2ex]
				As 	&	$g_j^D$	&	1.996343 	&	0.815597 	&	1.201699 	&	0.662240 	&	1.323779 	\\
				&	$\Delta g_j^Q$	&	0.009245 	&	$-$0.001698 	&	0.001859 	&	$-$0.003091 	&	0.002963 	\\
				&	$\Delta_{cor}$	&	$-$0.000012 	&	$-$0.000290 	&	$-$0.000122 	&	$-$0.000286 	&	0.000699 	\\
				&	$\Delta_{virt}$	&	$-$0.000092 	&	$-$0.000867 	&	0.000973 	&	$-$0.003714 	&	0.003683 	\\
				&	$g_j^{Total}$	&	2.005484 	&	0.812741 	&	1.204408 	&	0.655149 	&	1.331123 	\\
				&	Uncert.	&	0.000093 	&	0.000914 	&	0.000981 	&	0.003725 	&	0.003749 	\\ [+2ex]
				Se$^{+}$	&	$g_j^D$	&	1.993716 	&	0.827341 	&	1.200270 	&	0.665992 	&	1.312132 	\\
				&	$\Delta g_j^Q$	&	0.009221 	&	$-$0.003182 	&	$-$0.001273 	&	0.000849 	&	0.000529 	\\
				&	$\Delta_{cor}$	&	$-$0.000001 	&	$-$0.000541 	&	$-$0.000029 	&	$-$0.000002 	&	0.000569 	\\
				&	$\Delta_{virt}$	&	$-$0.000112 	&	0.000038 	&	0.000169 	&	$-$0.000439 	&	0.000331 	\\
				&	$g_j^{Total}$	&	2.002826 	&	0.823655 	&	1.199137 	&	0.666400 	&	1.313561 	\\
				&	Uncert.	&	0.000112 	&	0.000542 	&	0.000171 	&	0.000439 	&	0.000658 	\\ [+2ex]
				Br$^{2+}$	&	$g_j^D$	&	1.989802 	&	0.840605 	&	1.199930 	&	0.666474 	&	1.302406 	\\
				&	$\Delta g_j^Q$	&	0.009186 	&	$-$0.002934 	&	$-$0.001173 	&	0.000782 	&	0.000474 	\\
				&	$\Delta_{cor}$	&	0.000000 	&	$-$0.000613 	&	0.000001 	&	0.000000 	&	0.000610 	\\
				&	$\Delta_{virt}$	&	$-$0.000175 	&	0.000289 	&	0.000023 	&	$-$0.000033 	&	$-$0.000116 	\\
				&	$g_j^{Total}$	&	1.998813 	&	0.837346 	&	1.198780 	&	0.667223 	&	1.303374 	\\
				&	Uncert.	&	0.000175 	&	0.000678 	&	0.000023 	&	0.000033 	&	0.000621 	\\ \hline \hline
		\end{tabular}}
	\end{table*}
	
	\begin{table*}[btp]
		\caption{The $g_J$ factor for the $np^3$ $^4S_{3/2}$, $^2D_{3/2,5/2}$, and $^2P_{3/2,1/2}$ states ($n=5-6$). \label{gj_2}}
		{\setlength{\tabcolsep}{12pt}
			\begin{tabular}{ll rrrrr}\hline\hline \addlinespace[0.2cm]
				System	&		&	\multicolumn{1}{c}{$^4S_{3/2}$}	&	\multicolumn{1}{c}{$^2D_{3/2}$}	&	\multicolumn{1}{c}{$^2D_{5/2}$}	&	\multicolumn{1}{c}{$^2P_{1/2}$}	&	\multicolumn{1}{c}{$^2P_{3/2}$}	\\ \hline \addlinespace[0.1cm]
				Sb 	&	$g_j^D$	&	1.9717606 	&	0.8771228 	&	1.2000095 	&	0.6661579 	&	1.2847547 	\\
				&	$\Delta g_j^Q$	&	0.0090268 	&	$-$0.0022504 	&	$-$0.0009005 	&	0.0006006 	&	0.0003426 	\\
				&	$\Delta_{cor}$	&	0.0004467 	&	$-$0.0038865 	&	0.0001483 	&	$-$0.0008985 	&	0.0042532 	\\
				&	$\Delta_{virt}$	&	$-$0.0002223 	&	0.0002151 	&	0.0000096 	&	$-$0.0000616 	&	0.0001004 	\\
				&	$g_j^{Total}$	&	1.9810118 	&	0.8712010 	&	1.1992670 	&	0.6657984 	&	1.2894509 	\\
				&	Uncert.	&	0.0004990 	&	0.0038925 	&	0.0001486 	&	0.0009006 	&	0.0042543 	\\
				&	Exp. 1988\cite{Hassini-JOSA-1988}	&	1.967 $\ \ \ \ \ $	&	0.889 $\ \ \ \ \ $	&	1.205 $\ \ \ \ \ $	&	0.688 $\ \ \ \ \ $	&	1.277 $\ \ \ \ \ $	\\
				&	Exp. 2016\cite{Sobolewski-EPJD-2016}	&		&		&		&	0.676 $\ \ \ \ \ $	&	1.279 $\ \ \ \ \ $	\\ [+2ex]
				Te$^{+}$	&	$g_j^D$	&	1.9516918 	&	0.9082274 	&	1.1998492 	&	0.6664472 	&	1.2733024 	\\
				&	$\Delta g_j^Q$	&	0.0088466 	&	$-$0.0016687 	&	$-$0.0006675 	&	0.0004451 	&	0.0002440 	\\
				&	$\Delta_{cor}$	&	0.0006675 	&	$-$0.0036881 	&	$-$0.0000009 	&	$-$0.0000924 	&	0.0031347 	\\
				&	$\Delta_{virt}$	&	$-$0.0001875 	&	0.0002244 	&	$-$0.0000287 	&	$-$0.0000538 	&	0.0000438 	\\
				&	$g_j^{Total}$	&	1.9610183 	&	0.9030950 	&	1.1991521 	&	0.6667461 	&	1.2767249 	\\
				&	Uncert.	&	0.0006933 	&	0.0036949 	&	0.0000287 	&	0.0001069 	&	0.0031350 	\\
				&	NIST\cite{NIST}	&	1.93 $\ \ \ \ \ \ $	&		&		&		&	1.27 $\ \ \ \ \ \ $	\\ [+2ex]
				I$^{2+}$	&	$g_j^D$	&	1.9225889 	&	0.9454574 	&	1.1997966 	&	0.6664881 	&	1.2649503 	\\
				&	$\Delta g_j^Q$	&	0.0085860 	&	$-$0.0009960 	&	$-$0.0003984 	&	0.0002656 	&	0.0001413 	\\
				&	$\Delta_{cor}$	&	0.0007877 	&	$-$0.0033382 	&	$-$0.0000021 	&	$-$0.0000048 	&	0.0025506 	\\
				&	$\Delta_{virt}$	&	$-$0.0001833 	&	0.0001944 	&	$-$0.0000181 	&	$-$0.0000142 	&	0.0000279 	\\
				&	$g_j^{Total}$	&	1.9317793 	&	0.9413176 	&	1.1993780 	&	0.6667348 	&	1.2676701 	\\
				&	Uncert.	&	0.0008088 	&	0.0033438 	&	0.0000182 	&	0.0000150 	&	0.0025507 	\\ [+2ex]
				Bi 	&	$g_j^D$	&	1.6588424 	&	1.2148776 	&	1.2001881 	&	0.6664614 	&	1.2607733 	\\
				&	$\Delta g_j^Q$	&	0.0061725 	&	0.0025754 	&	0.0010320 	&	$-$0.0006883 	&	$-$0.0003591 	\\
				&	$\Delta_{cor}$	&	0.0010476 	&	$-$0.0006767 	&	$-$0.0022391 	&	$-$0.0009671 	&	0.0024275 	\\
				&	$\Delta_{virt}$	&	$-$0.0009516 	&	0.0012311 	&	$-$0.0001282 	&	$-$0.0000434 	&	$-$0.0004422 	\\
				&	$g_j^{Total}$	&	1.6651110 	&	1.2180073 	&	1.1988529 	&	0.6647626 	&	1.2623994 	\\
				&	Uncert.	&	0.0014153 	&	0.0014049 	&	0.0022428 	&	0.0009680 	&	0.0024674 	\\
				&	Exp. 1985\cite{George-JOSAB-1985}	&		&	1.225 $\ \ \ \ \ $	&	1.2 $\ \ \ \ \ \ \ $	&	0.667 $\ \ \ \ \ $	&		\\ [+2ex]
				Po$^{+}$	&	$g_j^D$	&	1.5865917 	&	1.2837211 	&	1.2001091 	&	0.6665329 	&	1.2635258 	\\
				&	$\Delta g_j^Q$	&	0.0054871 	&	0.0030615 	&	0.0012258 	&	$-$0.0008174 	&	$-$0.0004311 	\\
				&	$\Delta_{cor}$	&	0.0009360 	&	$-$0.0000332 	&	$-$0.0017831 	&	$-$0.0020087 	&	0.0027685 	\\
				&	$\Delta_{virt}$	&	0.0002473 	&	0.0004217 	&	$-$0.0000078 	&	0.0000247 	&	$-$0.0006887 	\\
				&	$g_j^{Total}$	&	1.5932622 	&	1.2871710 	&	1.1995440 	&	0.6637314 	&	1.2651745 	\\
				&	Uncert.	&	0.0009681 	&	0.0004230 	&	0.0017831 	&	0.0020089 	&	0.0028529 	\\ [+2ex]
				At$^{2+}$	&	$g_j^D$	&	1.5334809 	&	1.3322873 	&	1.1999486 	&	0.6664755 	&	1.2678098 	\\
				&	$\Delta g_j^Q$	&	0.0049864 	&	0.0032765 	&	0.0013114 	&	$-$0.0008746 	&	$-$0.0004686 	\\
				&	$\Delta_{cor}$	&	$-$0.0012440 	&	$-$0.0008953 	&	0.0014301 	&	0.0014944 	&	$-$0.0009308 	\\
				&	$\Delta_{virt}$	&	0.0000288 	&	$-$0.0077050 	&	0.0094668 	&	$-$0.0002049 	&	$-$0.0009643 	\\
				&	$g_j^{Total}$	&	1.5372522 	&	1.3269636 	&	1.2121569 	&	0.6668904 	&	1.2654461 	\\
				&	Uncert.	&	0.0012443 	&	0.0077568 	&	0.0095742 	&	0.0015084 	&	0.0013402 	\\ \hline \hline
		\end{tabular}}
	\end{table*}
	
	\begin{table*}[btp]
		\caption{Lifetimes $\tau$ (in s) for the $np^3$ $^2D_{3/2,5/2}$ and $^2P_{3/2,1/2}$ excited states ($n=3-6$). The numbers in brackets 
			are the uncertainties. \label{tau}}
		{\setlength{\tabcolsep}{16pt}
			\begin{tabular}{ll ll ll}\hline\hline \addlinespace[0.2cm]
				System	&	Method	&	$^2D_{3/2}$	&	$^2D_{5/2}$	&	$^2P_{1/2}$	&	$^2P_{3/2}$	\\ \hline \addlinespace[0.1cm]
				P 	&	MRCI	&	7170(165)	&	10800(316)	&	7.03(6)	&	5.29(3)	\\
				&	SCI 1982\cite{Mendoza-MNRAS-1982} 	&	7452 	&	11813 	&	6.04 	&	4.73 	\\
				&	MCDF 1999\cite{Fritzsche-APJ-1999} 	&	6711 	&	13545 	&	7.98 	&	6.01 	\\
				&	Emp. 1963\cite{Czyzak-MNRAS-1963} 	&	3378 	&	5152 	&	5.08 	&	3.38 	\\ [+2ex]
				S$^{+}$	&	MRCI	&	1070(1)	&	3510(45)	&	3.19(1)	&	1.88(1)	\\
				&	MCDF 2019\cite{Rynkun-AA-2019}	&	1060 	&	3780 	&	3.10 	&	1.86 	\\
				&	SCI 1982\cite{Mendoza-MNRAS-1982}	&	1134 	&	3840 	&	3.02 	&	1.86 	\\
				&	MCDF 1984\cite{Huang-ADNDT-1984}	&	1027 	&	2082 	&	2.62 	&	1.48 	\\
				&	MCDF 1999\cite{Fritzsche-APJ-1999}	&	980 	&	4091 	&	3.37 	&	1.96 	\\
				&	CI 1993\cite{Keenan-PS-1993}	&	810 	&	3496 	&	3.03 	&	1.70 	\\
				&	MCDF 2006\cite{Fischer-ADNDT-2006}	&	1378 	&	4419 	&	3.62 	&	2.22 	\\
				&	MCHF-BP 2005\cite{Irimia-PS-2005}	&	1461 	&	4936 	&	3.46 	&	2.16 	\\
				&	MCHF-BP 2010\cite{Tayal-APJSS-2010}	&	1580 	&	4550 	&	3.39 	&	2.14 	\\ [+2ex]
				Cl $^{2+}$	&	MRCI	&	190(1)	&	1290(10)	&	1.40(1)	&	0.69(1)	\\
				&	MCDF 2019\cite{Rynkun-AA-2019}	&	183 	&	1350 	&	1.41 	&	0.70 	\\
				&	MCDF 1999\cite{Fritzsche-APJ-1999}	&	197 	&	1504 	&	1.46 	&	0.71 	\\
				&	SCI 1982\cite{Mendoza-MNRAS-1982}	&	207 	&	1413 	&	1.41 	&	0.72 	\\
				&	MCDF 1984\cite{Huang-ADNDT-1984}	&	175 	&	784 	&	1.21 	&	0.59 	\\
				&	Emp. 1963\cite{Czyzak-MNRAS-1963}	&	141 	&	990 	&	1.20 	&	0.58 	\\ [+2ex]
				As 	&	MRCI	&	21(3)	&	248(4)	&	1.08(2)	&	0.47(1)	\\
				&	Emp. 1964\cite{Garstang-JRNBS-1964}	&	13 	&	177 	&	0.85 	&	0.55 	\\
				&	HFR 1986\cite{Biemont-PS-1986}	&	12 	&	153 	&	0.79 	&	0.35 	\\
				&	HXR 1986\cite{Biemont-PS-1986}	&	12 	&	143 	&	0.77 	&	0.34 	\\ [+2ex]
				Se$^{+}$	&	MRCI	&	4.14(2)	&	64(2)	&	0.34(1)	&	0.14(1)	\\
				&	HFR 1986\cite{Biemont-PS-1986}	&	2.86 	&	46 	&	0.28 	&	0.12 	\\
				&	HXR 1986\cite{Biemont-PS-1986}	&	3 	&	45 	&	0.27 	&	0.12 	\\ [+2ex]
				Br$^{2+}$	&	MRCI	&	1.24(1)	&	16(1)	&	0.14(3)	&	0.05(1)	\\
				&	HFR 1986\cite{Biemont-PS-1986}	&	0.87 	&	15 	&	0.12 	&	0.05 	\\
				&	HXR 1986\cite{Biemont-PS-1986}	&	0.87 	&	15 	&	0.12 	&	0.05 	\\ [+2ex]
				Sb 	&	MRCI	&	1.40(7)	&	13(1)	&	0.27(1)	&	0.10(1)	\\
				&	Emp. 1964\cite{Garstang-JRNBS-1964}	&	0.90 	&	10 	&	0.21 	&	0.09 	\\
				&	HFR 1995\cite{Biemont-AASS-1995}	&	0.90 	&	9.1 	&	0.20 	&	0.09 	\\ [+2ex]
				Te$^{+}$	&	MRCI	&	0.37(1)	&	3.13(1)	&	0.09(3)	&	0.034(1)	\\
				&	HFR 1995\cite{Biemont-AASS-1995}	&	0.27 	&	2.40 	&	0.07 	&	0.031 	\\ [+2ex]
				I$^{2+}$	&	MRCI	&	0.19(6)	&	1.29(1)	&	0.043(6)	&	0.015(1)	\\
				&	HFR 1995\cite{Biemont-AASS-1995}	&	0.106 	&	0.797 	&	0.153 	&	0.014 	\\ [+2ex]
				Bi 	&	MRCI	&	0.036(1)	&	0.151(1)	&	0.019(1)	&	0.0058(1)	\\
				&	Emp. 1964\cite{Garstang-JRNBS-1964}	&	0.032 	&	0.119 	&	0.016 	&	0.0058 	\\
				&	HFR 1996\cite{Biemont-PS-1996}	&	0.025 	&	0.087 	&	0.0150 	&	0.0062 	\\ [+2ex]
				Po$^{+}$	&	MRCI	&	0.0107(1)	&	0.043(1)	&	0.0060(1)	&	0.0021(3)	\\
				&	HFR 1996\cite{Biemont-PS-1996}	&	0.001 	&	0.037 	&	0.0059 	&	0.0024 	\\ [+2ex]
				At$^{2+}$	&	MRCI	&	0.0040(1)	&	0.0164(1)	&	0.0026(2)	&	0.0009(2)	\\
				&	HFR 1996\cite{Biemont-PS-1996}	&	0.0038 	&	0.0149 	&	0.0026 	&	0.0010 	\\ \hline \hline
		\end{tabular}}
	\end{table*}
	
	\begin{table*}[btp]
		\caption{Magnetic dipole hyperfine structure constants $A$ (in MHz) for the $np^3$ $^4S_{3/2}$, $^2D_{3/2,5/2}$ and $^2P_{3/2,1/2}$ states ($n=3-4$).
			The numbers in brackets are the uncertainties that are estimated as the root mean square of $\Delta_{cor}$ and $\Delta_{virt}$. The nuclear 
			magnetic moment $\mu_N$ value is taken from Ref. \cite{Stone-ADNDT-2005}. \label{HFSA-1}}
		{\setlength{\tabcolsep}{8pt}
			\begin{tabular}{ll ll rrrrr}\hline\hline \addlinespace[0.2cm]
				System	&	I	&	$\mu_N$	&	Model	&	$^4S_{3/2}$	&	$^2D_{3/2}$	&	$^2D_{5/2}$	&	$^2P_{1/2}$	&	$^2P_{3/2}$	 \\ \hline \addlinespace[0.1cm]
				$^{31}$P	&	1/2	&	1.1316 	&	Core10SV2SDV3SDT/aae4z	&	93.84 	&	430.13 	&	739.80 	&	1998.06 	&	430.69 	\\
				&		&		&	$\Delta_{cor}$	&	1.70 	&	0.74 	&	2.71 	&	8.88 	&	2.16 	\\
				&		&		&	$\Delta_{virt}$	&	19.45 	&	9.86 	&	32.14 	&	77.86 	&	20.08 	\\
				&		&		&	Final	&	115(20)	&	441(10)	&	775(32)	&	2085(78)	&	453(20)	\\ [+2ex]
				$^{33}$S$^+$	&	3/2	&	0.6438 	&	Core10SV2SDV3SDT/aae4z	&	38.21 	&	141.77 	&	244.80 	&	652.97 	&	143.59 	\\
				&		&		&	$\Delta_{cor}$	&	0.45 	&	0.54 	&	0.87 	&	2.74 	&	0.87 	\\
				&		&		&	$\Delta_{virt}$	&	5.23 	&	2.82 	&	9.03 	&	22.00 	&	5.56 	\\
				&		&		&	Final	&	44(5)	&	145(3)	&	255(9)	&	675(22)	&	150(6)	\\ [+2ex]
				$^{35}$Cl$^{2+}$	&	3/2	&	0.8219 	&	Core10SV2SDV3SDT/aae4z	&	76.63 	&	280.75 	&	483.47 	&	1288.29 	&	283.48 	\\
				&		&		&	$\Delta_{cor}$	&	0.72 	&	1.13 	&	1.79 	&	5.18 	&	1.62 	\\
				&		&		&	$\Delta_{virt}$	&	8.26 	&	4.98 	&	15.41 	&	38.23 	&	9.30 	\\
				&		&		&	Final	&	86(8)	&	287(5)	&	501(16)	&	1332(39)	&	294(9)	\\ [+2ex]
				$^{75}$As	&	3/2	&	1.4395 	&	Core18SV2SDV3SDT/aae4z	&	$-$75.69 	&	383.16 	&	646.22 	&	1959.84 	&	326.34 	\\
				&		&		&	$\Delta_{cor}$	&	4.80 	&	1.72 	&	5.80 	&	12.13 	&	3.80 	\\
				&		&		&	$\Delta_{virt}$	&	$-$32.70 	&	21.13 	&	8.88 	&	59.84 	&	$-$9.00 	\\
				&		&		&	Final	&	$-$104(33)	&	406(21)	&	661(11)	&	2032(61)	&	321(10)	\\
				&		&		&	Exp. 1987\cite{Bouazza-ZPD-1987}	&		&	480(9)	&	725(26)	&	1909(14)	&	359(59)	\\ [+2ex]
				$^{77}$Se$^+$	&	1/2	&	0.5355 	&	Core18SV2SDV3SDT/aae4z	&	$-$120.37 	&	660.58 	&	1138.37 	&	3430.49 	&	583.02 	\\
				&		&		&	$\Delta_{cor}$	&	6.79 	&	1.72 	&	7.19 	&	16.17 	&	4.80 	\\
				&		&		&	$\Delta_{virt}$	&	$-$39.86 	&	23.63 	&	16.04 	&	82.46 	&	$-$4.20 	\\
				&		&		&	Final	&	$-$153(40)	&	686(24)	&	1162(18)	&	3529(84)	&	584(6)	\\ [+2ex]
				$^{79}$Br$^{2+}$	&	3/2	&	2.1064 	&	Core18SV2SDV3SDT/aae4z	&	$-$114.77 	&	1208.43 	&	2160.71 	&	6411.48 	&	1134.43 	\\
				&		&		&	$\Delta_{cor}$	&	9.47 	&	3.01 	&	9.68 	&	22.21 	&	6.70 	\\
				&		&		&	$\Delta_{virt}$	&	$-$58.46 	&	34.09 	&	26.58 	&	131.64 	&	$-$4.05 	\\
				&		&		&	Final	&	$-$164(59)	&	1245(34)	&	2197(28)	&	6565(134)	&	1137(8)	\\ \hline \hline
		\end{tabular}}
	\end{table*}
	
	\begin{table*}[btp]
		\caption{Magnetic dipole hyperfine structure constants $A$ (in MHz) for the $np^3$ $^4S_{3/2}$, $^2D_{3/2,5/2}$ and $^2P_{3/2,1/2}$ states ($n=5-6$). 
			The numbers in brackets are the uncertainties that are estimated as the root mean square of $\Delta_{cor}$ and $\Delta_{virt}$. The $\mu_N$ value is 
			taken from Ref. \cite{Stone-ADNDT-2005}. \label{HFSA-2}}
		{\setlength{\tabcolsep}{6pt}
			\begin{tabular}{ll ll rrrrr}\hline\hline \addlinespace[0.2cm]
				System	&	I	&	$\mu_N$	&	Model	&	$^4S_{3/2}$	&	$^2D_{3/2}$	&	$^2D_{5/2}$	&	$^2P_{1/2}$	&	$^2P_{3/2}$	 \\ \hline \addlinespace[0.1cm]
				$^{121}$Sb	&	5/2	&	3.3592 	&	Core18SV2SDV3SDT/aae4z	&	$-$401.41 	&	613.86 	&	1374.65 	&	4688.08 	&	597.11 	\\
				&		&		&	$\Delta_{cor}$	&	45.76 	&	$-$6.58 	&	$-$6.81 	&	$-$61.56 	&	$-$2.89 	\\
				&		&		&	$\Delta_{virt}$	&	$-$57.78 	&	30.70 	&	14.32 	&	89.87 	&	$-$5.09 	\\
				&		&		&	Final	&	$-$413(74)	&	638(31)	&	1382(16)	&	4716(109)	&	589(6)	\\
				&		&		&	Exp. 1978\cite{Buchholz-ZPA-1978}	&		&	563(15)	&	1465(89)	&	4949(89)	&	673(42)	\\
				&		&		&	Exp. 1988\cite{Hassini-JOSA-1988}	&	$-$305.79(21)	&	565(7)	&	1468.98	&	4907.60	&	680(1)	\\ [+2ex]
				$^{125}$Te$^+$	&	1/2	&	$-$0.8871 	&	Core18SV2SDV3SDT/aae4z	&	375.59 	&	$-$1098.16 	&	$-$2778.98 	&	$-$9108.30 	&	$-$1288.89 	\\
				&		&		&	$\Delta_{cor}$	&	$-$69.63 	&	13.38 	&	9.89 	&	95.30 	&	6.21 	\\
				&		&		&	$\Delta_{virt}$	&	79.61 	&	$-$38.44 	&	$-$27.63 	&	$-$161.73 	&	11.18 	\\
				&		&		&	Final	&	386(106)	&	$-$1123(41)	&	$-$2797(30)	&	$-$9175(188)	&	$-$1272(13)	\\ [+2ex]
				$^{127}$I$^{2+}$	&	5/2	&	2.8080 	&	Core18SV2SDV3SDT/aae4z	&	$-$312.73 	&	842.83 	&	2440.26 	&	7848.56 	&	1167.56 	\\
				&		&		&	$\Delta_{cor}$	&	43.14 	&	$-$7.88 	&	$-$6.80 	&	$-$62.05 	&	$-$1.55 	\\
				&		&		&	$\Delta_{virt}$	&	$-$59.11 	&	24.51 	&	21.58 	&	127.15 	&	$-$8.04 	\\
				&		&		&	Final	&	$-$329(73)	&	859(26)	&	2455(23)	&	7914(141)	&	1158(8)	\\ [+2ex]
				$^{209}$Bi	&	9/2	&	4.1100 	&	Core18SV2SDV3SDT/aae4z	&	$-$502.25 	&	$-$1118.46 	&	2460.47 	&	11005.00 	&	456.13 	\\
				&		&		&	$\Delta_{cor}$	&	23.52 	&	3.57 	&	15.80 	&	$-$63.38 	&	$-$42.43 	\\
				&		&		&	$\Delta_{virt}$	&	47.32 	&	34.36 	&	54.16 	&	83.08 	&	17.39 	\\
				&		&		&	Final	&	$-$431(53)	&	$-$1081(35)	&	2530(56)	&	11025(105)	&	431(46)	\\
				&		&		&	SE 2021\cite{Wilman-JQSRT-2021}	&	$-$448.312	&	$-$1225.52	&	2525.71 	&	11252.2	&	484.855	\\
				&		&		&	Exp. 2021\cite{Wilman-JQSRT-2021}	&	$-$446.937	&	$-$1231.02	&	2502.86	&	11260.2	&	491.028	\\
				&		&		&	Exp. 2007\cite{Wasowicz-PS-2007}	&	$-$447.52	&		&	2508.36	&	11272.198	&		\\
				&		&		&	Exp. 2000\cite{Pearson-JPG-2000}	&	$-$446.97	&		&		&		&		\\ [+2ex]
				$^{210}$Po$^+$	&	9/2	&	$-$0.3800 	&	Core18SV2SDV3SDT/aae4z	&	$-$16.95 	&	$-$407.26 	&	884.31 	&	3772.21 	&	198.70 	\\
				&		&		&	$\Delta_{cor}$	&	2.45 	&	$-$4.25 	&	5.39 	&	$-$0.87 	&	$-$7.65 	\\
				&		&		&	$\Delta_{virt}$	&	5.82 	&	1.81 	&	14.08 	&	40.07 	&	0.93 	\\
				&		&		&	Final	&	$-$9(6)	&	$-$410(5)	&	904(15)	&	3811(40)	&	192(8)	\\ [+2ex]
				$^{210}$At$^{2+}$	&	5	&	4.7400 	&	Core18SV2SDV3SDT/aae4z	&	395.85 	&	$-$2505.46 	&	5381.35 	&	22387.28 	&	1321.98 	\\
				&		&		&	$\Delta_{cor}$	&	10.83 	&	12.65 	&	$-$12.09 	&	49.85 	&	$-$19.30 	\\
				&		&		&	$\Delta_{virt}$	&	5.95 	&	238.11 	&	$-$198.45 	&	260.64 	&	5.05 	\\
				&		&		&	Final	&	413(12)	&	$-$2255(238)	&	5171(199)	&	22698(265)	&	1308(20)	\\ \hline \hline
		\end{tabular}}
	\end{table*}
	
	\begin{table*}[btp]
		\caption{Electric quadrupole hyperfine structure constants $B$ (in MHz) for the $np^3$ $^4S_{3/2}$, $^2D_{3/2,5/2}$ and $^2P_{3/2}$ levels of
			$np^3$ ($n=3-6$). The numbers in brackets are the uncertainties that are estimated as the root mean square of $\Delta_{cor}$ and $\Delta_{virt}$. The
			value of the nuclear electric quadruple moment $Q_s$ is taken from Ref. \cite{Stone-ADNDT-2005}. \label{HFSB-1}}
		{\setlength{\tabcolsep}{14pt}
			\begin{tabular}{lll rrrr}\hline\hline \addlinespace[0.2cm]
				System	&	$Q_s$	&	Model	&	$^4S_{3/2}$	&	$^2D_{3/2}$	&	$^2D_{5/2}$	&	$^2P_{3/2}$	\\ \hline \addlinespace[0.1cm]
				$^{33}$S$^+$	&	$-$0.678	&	Core10SV2SDV3SDT/aae4z	&	0.060 	&	$-$43.178 	&	14.011 	&	43.027 	\\
				&		&	$\Delta_{cor}$	&	0.012 	&	0.192 	&	0.103 	&	$-$0.832 	\\
				&		&	$\Delta_{virt}$	&	0.004 	&	$-$1.726 	&	$-$0.059 	&	1.388 	\\
				&		&	Final	&	0.08(1)	&	$-$45(2)	&	14.1(2)	&	44(2)	\\ [+2ex]
				$^{35}$Cl$^{2+}$	&	0.085	&	Core10SV2SDV3SDT/aae4z	&	$-$0.015 	&	12.887 	&	$-$2.012 	&	$-$13.008 	\\
				&		&	$\Delta_{cor}$	&	0.001 	&	0.002 	&	$-$0.013 	&	0.069 	\\
				&		&	$\Delta_{virt}$	&	$-$0.001 	&	0.432 	&	0.050 	&	$-$0.397 	\\
				&		&	Final	&	$-$0.015(1)	&	13.3(4)	&	$-$1.98(5)	&	$-$13.3(4)	\\ [+2ex]
				$^{75}$As	&	0.3	&	Core18SV2SDV3SDT/aae4z	&	$-$0.284 	&	93.343 	&	$-$7.654 	&	$-$92.503 	\\
				&		&	$\Delta_{cor}$	&	$-$0.015 	&	$-$0.217 	&	$-$0.094 	&	1.537 	\\
				&		&	$\Delta_{virt}$	&	0.074 	&	5.123 	&	2.365 	&	$-$6.895 	\\
				&		&	Final	&	$-$0.22(7)	&	98(5)	&	$-$5.4(24)	&	$-$98(7)	\\ [+2ex]
				$^{77}$Se$^+$	&	0.535	&	Core18SV2SDV3SDT/aae4z	&	$-$0.351 	&	334.254 	&	$-$17.122 	&	$-$333.372 	\\
				&		&	$\Delta_{cor}$	&	$-$0.004 	&	$-$1.077 	&	$-$0.207 	&	3.008 	\\
				&		&	$\Delta_{virt}$	&	0.094 	&	10.243 	&	2.712 	&	$-$10.939 	\\
				&		&	Final	&	$-$0.26(9)	&	343(10)	&	$-$15(3)	&	$-$341(11)	\\ [+2ex]
				$^{79}$Br$^{2+}$	&	0.318	&	Core18SV2SDV3SDT/aae4z	&	0.907 	&	337.775 	&	$-$12.158 	&	$-$338.564 	\\
				&		&	$\Delta_{cor}$	&	0.027 	&	$-$0.690 	&	0.096 	&	1.812 	\\
				&		&	$\Delta_{virt}$	&	0.058 	&	7.905 	&	1.565 	&	$-$8.448 	\\
				&		&	Final	&	0.99(6)	&	345(8)	&	$-$11(2)	&	$-$345(9)	\\ [+2ex]
				$^{121}$Sb	&	$-$0.36	&	Core18SV2SDV3SDT/aae4z	&	$-$1.63 	&	$-$316.31 	&	9.01 	&	320.59 	\\
				&		&	$\Delta_{cor}$	&	$-$0.19 	&	9.08 	&	$-$0.47 	&	$-$10.58 	\\
				&		&	$\Delta_{virt}$	&	$-$0.14 	&	$-$9.90 	&	$-$4.32 	&	10.64 	\\
				&		&	Final	&	$-$1.96(23)	&	$-$317(13)	&	4(4)	&	321(15)	\\
				&		&	Exp. 1978\cite{Buchholz-ZPA-1978}	&		&	$-$360(89)	&		&	414(36)	\\
				&		&	Exp. 1988\cite{Hassini-JOSA-1988}	&		&	$-$561(6)	&		&	438(30)	\\ [+2ex]
				$^{125}$Te$^+$	&	0.58	&	Core18SV2SDV3SDT/aae4z	&	26.12 	&	816.31 	&	$-$18.76 	&	$-$844.87 	\\
				&		&	$\Delta_{cor}$	&	$-$0.28 	&	$-$17.63 	&	0.07 	&	20.14 	\\
				&		&	$\Delta_{virt}$	&	0.67 	&	17.79 	&	5.39 	&	$-$18.99 	\\
				&		&	Final	&	26.5(7)	&	816(25)	&	$-$13(5)	&	$-$844(28)	\\ [+2ex]
				$^{127}$I$^{2+}$	&	0.72	&	Core18SV2SDV3SDT/aae4z	&	104.35 	&	1411.23 	&	$-$28.14 	&	$-$1516.29 	\\
				&		&	$\Delta_{cor}$	&	$-$0.97 	&	$-$22.26 	&	$-$0.17 	&	29.08 	\\
				&		&	$\Delta_{virt}$	&	2.18 	&	24.69 	&	6.16 	&	$-$27.20 	\\
				&		&	Final	&	106(2)	&	1414(33)	&	$-$22(6)	&	$-$1514(40)	\\ [+2ex]
				$^{209}$Bi	&	$-$0.516	&	Core18SV2SDV3SDT/aae4z	&	$-$332.9 	&	$-$787.7 	&	18.6 	&	1167.7 	\\
				&		&	$\Delta_{cor}$	&	17.5 	&	12.0 	&	4.1 	&	26.9 	\\
				&		&	$\Delta_{virt}$	&	$-$10.9 	&	$-$31.8 	&	$-$5.3 	&	32.4 	\\
				&		&	Final	&	$-$326.3 	&	$-$807.5 	&	17.4 	&	1227.0 	\\
				&		&	Exp. 1985\cite{George-JOSAB-1985}	&	$-$324(21)	&	$-$609(34)	&	57(7)	&	1025(42)	\\
				&		&	Exp. 2021\cite{Wilman-JQSRT-2021}	&	$-$305.07	&	$-$652.50	&	23	&	978.64	\\
				&		&	Exp. 2007\cite{Wasowicz-PS-2007}	&	$-$305.47	&		&	38.97	&		\\
				&		&	Exp. 2000\cite{Pearson-JPG-2000}	&	$-$304.30	&		&		&		\\ [+2ex]
				$^{211}$Po$^+$	&	$-$0.57	&	Core18SV2SDV3SDT/aae4z	&	$-$779.6 	&	$-$1051.7 	&	25.0 	&	1876.5 	\\
				&		&	$\Delta_{cor}$	&	25.7 	&	18.9 	&	12.9 	&	20.9 	\\
				&		&	$\Delta_{virt}$	&	$-$13.0 	&	$-$28.1 	&	$-$20.6 	&	37.3 	\\
				&		&	Final	&	$-$767(29)	&	$-$1061(34)	&	17(24)	&	1935(43)	\\ [+2ex]
				$^{210}$At$^{2+}$	&	0.68	&	Core18SV2SDV3SDT/aae4z	&	1504.4 	&	1399.6 	&	$-$37.1 	&	$-$2955.7 	\\
				&		&	$\Delta_{cor}$	&	40.7 	&	38.4 	&	44.3 	&	4.7 	\\
				&		&	$\Delta_{virt}$	&	17.1 	&	37.1 	&	5.9 	&	$-$52.5 	\\
				&		&	Final	&	1562(44)	&	1475(53)	&	13(45)	&	$-$3003(53)	\\ \hline \hline
		\end{tabular}}
	\end{table*}

	\section{Results and Discussion}
	
	The $p^3$ configuration has five fine structure splitting states, the ground state $^4S_{3/2}$, and the excited $^2D_{3/2,5/2}$ and $^2P_{1/2,3/2}$ 
	states. The excitation energies of the $^2D_{3/2,5/2}$, and $^2P_{1/2,3/2}$ states that are calculated by using the DC and RECP with different 
	types of basis sets are tabulated from Table \ref{tab:dyall} to Table \ref{np-lECP-cc} for the $n(=3-6)p^3$ configurations. The superheavy ion with the
	$7p^3$ configuration has many states with $^2P_{1/2,3/2}$ levels, apart from the other excited states as $7p^28s$, $7p^28p$, and $7p^27d$ whose energies
	are given in Table \ref{7p-EE}. We compare these values with the experimental values quoted in the National Institute of Standards and Technology (NIST) 
	database \cite{NIST}. Comparison of the calculated energies using both the DC and RECP Hamiltonians for each type of basis sets to understand trends 
	of correlation effects and accuracy of the results. The values obtained using the dyall.aaeXz basis set in the MRCI method with the DC Hamiltonian are 
	given in Table \ref{tab:dyall}. Convergence in the results are verified by performing calculations with $X$ = 2, 3, and 4 basis sets in the 
	Core10SV2SDV3SDT model approximation. Then, we extraploate the $E_{\rm{CBS}}$ values using Eq. (\ref{eq:dyallcbs}). We also estimate the $\Delta_{cor}$ 
	contributions using the Core10SDV2SDTV3SDTQ and Core18SV2SDV3SDT model calculations with the $X$ = 2 basis set. Differences in the results from the 
	Core10SV2SDV3SDT calculations with the virtual spinor cutoff values at 10 a.u. and  20 a.u. indicate about order of magnitudes of the $\Delta_{virt}$ 
	contributions. The net results listed under `Final' in the above table are taken as $E_{\rm{CBS}}$ along with the $\Delta_{cor}$ and $\Delta_{virt}$
	contributions. Uncertainties to these quantitues are mentioned as  `Uncert.' in the above table.
	
	From Table \ref{tab:dyall}, it can be observed that our calculations for P to Bi show good agreement with the NIST data. There are about 1\% descrepancies
	of the calculated energy values for the $^2D_{3/2,5/2}$ states with the experimental values, while they are about 2-6\% for the $^2P_{1/2,3/2}$ states. 
	From our analysis we found that inclusion correlation effections due to the inner core orbitals, frozen in the Core10SV2SDV3SDT model, or higher level 
	excitation configurations through the MRCI method do not improve the calculations substantially. However, we observe that these calculations are very 
	sensitive to the choice of basis functions. The above energy values were obtained using the dyall.aae4z basis set, the $\Delta_{basis}$ corrections 
	estimated due to increasing basis size are seen to be non-negligible. It implies that the descrepancies seen above for the calculated energies are 
	mainly due to use of the finite-size basis functions used in the MRCI calculations. We have also given energies of the Po$^+$ and At$^{2+}$ ions in 
	Table \ref{tab:dyall}, which has ground state configurations as $6p^3$. We could not find any experimental data for these ions to compare with our 
	calculations. Based on the comparsion of our calculated data for other neutral, single ionized and doubly ionized atoms with the $n(=3-5)p^3$ ground 
	state configurations, we can infer uncertainties to these calculations within 3\%.
	
	We present again the excitation energies of some of the afomentioned systems with $3p^3$ and $4p^3$ configurations in Table \ref{AEperterson} obtained 
	using the aug-cc-pVXZ basis sets in the MRCI method. It should be note that quality of the aug-cc-pVXZ basis set is better than the dyall.aaeXz basis 
	set, therefore they are more suitable to be used for the carrying calculations with the DC Hamiltonian in the MRCI method. They are able to include 
	contributions from more number of virtual orbitals. It increases up to 7 $\xi$ for systems with $3p$ valence orbitals and 5 $\xi$ for systems with $4p$ 
	valence orbitals. This may be the reason for which deviations of the MRCI calculations using the DC Hamiltonian from the NIST data compared to the 
	previously mentioned values obtained using the dyall.aaeXz basis set. Thus, it can be assumed that the aug-cc-pVXZ basis set are the better choice 
	over the dyall.aaeXz basis set for the light P-group elements from the third and forth rows of the periodic table.
	
	We also investigate how reliably the RECP Hamiltonian gives the energies when used instead of the DC Hamiltonian at the similar level of calculations. 
	For this purpose, we have considered the combined small-core PPs with the aug-cc-pVXZ-pp basis in the MRCI method set and the results are listed in 
	Table \ref{np-sECP-cc}. The use of the small-core PPs speeds up the MRCI computation significantly. Beside, we find that the contraction of the 
	aug-cc-pVXZ-pp basis set can include a larger number of virtual orbitals in the calculation using the MRCI method. We first carry out these calculations 
	under the Core18SV2SDV3SDT model approximation. These calculations with $X$ = 3, 4, and 5$\xi$ basis set show excellent convergence of the results and 
	are quoted as $E_{\rm{CBS}}$. We then improve the calculations by correlating electrons from more core orbitals under the core18SDV2SDTV3SDTQ model 
	appriximation and considering aug-cc-pV3Z-pp basis set. Corrections estimated in this approach are listed under $\Delta_{cor}$ in the above table.
	We consider the total value as the sum of $E_{\rm{CBS}}$ and $\Delta_{cor}$. Uncertainties to these calculations are estimated as the root mean square
	of contributions from $\Delta_{basis}$ and $\Delta_{cor}$, where $\Delta_{basis}$ are determined from the convergence of the results with respect 
	to the size of the basis functions. Comparison of these results with the NIST data suggest that the RECP results are reasonably good. For the 
	$^2P_{1/2,3/2}$ states, teh calculated excited energies are coming out to be around 2-4\% accuracy, which are slightly better than that were obtained 
	using the DC Hamiltonian. The RECP results for the Po$^+$ and At$^{2+}$ ions are comparable with the values obtained using the DC Hamiltonian.
	
	In the case of large core RECP approximation, we have conducted the V5SDTQ calculation with the aug-cc-pVXZ-pp basis sets for $X$ = 3, 4, and 5$\xi$. 
	The estimated values are listed in Table \ref{np-lECP-cc}. Comparing these values with the NIST data, we find that the energies almost agree 
	with the NIST data except for the first excited state, $^2D_{3/2}$. In some cases, discrepancies from these calculations lie within 5-12\%. This suggests
	that considering large core in the RECP Hamiltonian is not appropriate to obtain accurate results for the systems with $n(=4-6)p^3$ configurations. Moreover, the large-core PPs 
	results for Po$^+$ and At$^{2+}$ show unusually underestimated results to the energies compared to the values obtained for AE the small-core PP 
	calculations using the RECP Hamiltonian as well the results obtained using the DC Hamiltonian. 
	
	The above exercises for estimating energies accurately in the medium heavy atomic systems is meant to understand roles of various correlation 
	effects for pursuing accurate calculations of energies in the superheavy elements. In Table \ref{7p-EE}, we present the calculated energies of Mc, 
	Lv$^+$, and Ts$^{2+}$ and compare these values with the available literature data \cite{Dzuba-HI-2016}. Owing to the short-lived nature of the Mc, Lv, 
	and Ts isotopes, carrying out measurements of any spectroscopic properties of superheavy atoms and their ions are extremely challenging. 
	
	One particular aspect on which one would be very careful for carrying out accurate calculations of properties in the superheave elements is the 
	adequate choice of a sufficiently large basis set. Currently, the dyall.aae4z basis set is available for performing large relativistic calculations 
	using the DC Hamiltonian. Besides, we adopt the doubly-density (DD)-intensified procedure \cite{Hubert-2022} on the dyall.aae4z basis set, which makes
	the number of the $s$ and $p$ primary functions double. Then, we carry out the calculations in the Core10SV2SDV3SDT model using the DC Hamiltonian 
	in the MRCI method. We predict excitation energies of the states with the $7p^3$, $7p^28s$, $7p^28p$, and $7p^27d$ configurations from these 
	calculations. We find results from both the dyall.aae4z and DD-intensified dyall.aae4z basis sets are agreeing to each other. We also try to 
	improve the calculations using the Core10SDV2SDTV3SDTQ model and the dyall.aae4z basis set. In this case, we have truncated the virtual orbitals 
	considering energy cut-off at 2.0 a.u.. These corrections are listed as $\Delta_{cor}$ in Table \ref{7p-EE}. Thus, the final calculated values 
	follows as the sum of contributions from the DD-intensified dyall.aae4z basis set and $\Delta_{cor}$. Similarly, uncertainties to the final values 
	are estimated as the root mean square of $\Delta_{DD}$ and $\Delta_{cor}$. When compared our calculations for Mc with that are reported in
	Ref. \cite{Dzuba-HI-2016}, we find excellent agreement among the results from both the calculations. Thus, we assume that our estimated values for the 
	Lv$^+$, and Ts$^{2+}$ ions are reliable and can be used for future applications. 
	
	For completeness, we have also performed calculations of energies of the considered syperheavy systems using the RECP Hamiltonian in the MRCI 
	method. First we used the QZVP basis set to verify the results and found large differences between the results obtained using the RECP and DC 
	Hamiltonians. Then, we performed the calculations by replacing the QZVP basis set by the dyall.aae4z basis set. This showed very good agreement with 
	the results obtained using the DC Hamiltonian. So large differences in the results the former case can be attributed to the quality of the 
	QZVP basis set, which may be because of the small size of the QZVP basis set, which is not sufficient to produce the excited energies of the $7p$ 
	states. It is worth mentioning here is that previous studies on calculations of ionization potentials of these elements using the RECP Hamiltonian 
	and QZVP basis set showed reasonably accurate results \cite{Hangele-JCP-2012,Hangele-JCP-2013}. This implies that correlation behavior in the 
	determination of ionization potentials and excitation energies are different in Mc, Lv$^+$, and Ts$^{2+}$.
	
	After analyzing energies, we discuss below various spectroscopic properties of the undertaken atomic systems. We shall consider the same basis 
	sets for producing the properties that gave energies with reasonable accuracy. Since the properties under investigation have different radial 
	dependeny, they are expected to show different trends of electron correlation effects. Thus, we discuss these properties one after another.

	\subsection{Land\'{e} $g_J$ factor}
	
	The $g_J$ factors for the calculated states with $n(=3-6)p^3$ configurations of the considered systems are determined using the MRCI wave functions 
	obtained with the DC Hamiltonian. These values are listed in Tables \ref{gj_1} and \ref{gj_2}. We find that deciding accuracy of these quantities 
	are more challenging compared to the calculated energies. To ensure accurate determination of the $g_J$ factors of considered systems, we have 
	used the largest basis set and include correlation effects due to inner core orbitals. We employ the dyall.aae4z basis set in the Core18SV2SDV3SDT 
	and Core10SV2SDV3SDT model for the DC Hamiltonian for calculating the $g_j^D$ and $\Delta g_j^Q$ factors of the systems with 
	the $3p^3$ configurations. Further corrections are added through the higher level excitations in the Core10SDV2SDTV3SDTQ model and 
	using the dyall.aae2z basis set. This correction is listed as $\Delta_{cor}$ and corrections from the high-lying virtual orbitals are estimated 
	as $\Delta_{virt}$. In the end, the final value, $g_j^{Total}$, is determined by 
	\begin{equation}
		g_j^{Total}=g_j^D+\Delta g_j^Q+\Delta_{cor}+\Delta_{virt}.
	\end{equation}
	Also, we estimate the net uncertainty to the total value as taking the root mean square contributions of $\Delta_{cor}$ and $\Delta_{virt}$. We 
	find that uncertainties are appearing mostly either in the fourth or fifth decimal places for the lighter systems while at the third decimal places for
	the heavier elements with $n=5-6$. This much precision
	is reasonable enough at the moment to compare our results with the available experimental data as the measured values are precise only up to the 
	second decimal places. Nonetheless, the reported $g_J$ factors will be very interesting to studying properties of the fine-structure splitting 
	of the considered elements and guide the future experiments to improve accuracy of the measured data.
	
	\subsection{Lifetimes of excited states}
	
	Table \ref{tau} presents the calculated $\tau$ values for the $^2D_{3/2,5/2}$ and $^2P_{3/2,1/2}$ excited states with $n(=3-6)p^3$ configuration, 
	obtained using the same computational strategies that were followed to estimate the $g_J$ factors accurately. The uncertainties in the evaluated $\tau$
	values are determined by combining the $\Delta_{cor}$ and $\Delta_{virt}$ corrections that are obtained due to both the calculated M1 and E2 transition 
	matrix elements. We have compared our results with the available literature data and observed overall good agreement. One may notice from the above 
	table that there are significant discrepancies among various experimental $\tau$ values of the $^2D_{3/2,5/2}$ and $^2P_{3/2,1/2}$ states of P, 
	S$^+$, and Cl$^{2+}$ atoms reported in the literature. For example, for the case of S$^+$, the well-known discrepancy between theoretical and 
	experimental data for very long-lived states has been discussed using the data from the laboratory astrophysics \cite{Trabert-atoms-2020}. Our results 
	are consistent with the semi-empirical configuration interaction (SCI) data reported in Ref. \cite{Mendoza-MNRAS-1982} and the multiconfiguration 
	Dirac-Fock (MCDF) data reported in Refs. \cite{Huang-ADNDT-1984,Fritzsche-APJ-1999,Rynkun-AA-2019} over the years, whereas disagree with the CI data 
	that of Ref. \cite{Keenan-PS-1993}, the MCDF data of Ref. \cite{Fischer-ADNDT-2006}, and the MCHF-BP results from Refs. \cite{Irimia-PS-2005,Tayal-APJSS-2010}.
	
	For the As neural atom and the Se$^+$ and Br$^{2+}$ ions that have the $4p^3$ configuration as well as the Sb neutral atom and the Te$^+$ and I$^{2+}$ 
	ions having the $5p^3$ configuration, one can find very old studies of $\tau$ values using the emperical approaches \cite{Garstang-JRNBS-1964},
	relativistic Hartree-plus-statistical-exchange (HXR) and Hartree-Fock (HFR) methods \cite{George-JOSAB-1985,Biemont-AASS-1995}. Our $\tau$ values
	show differences in the results with a factor about 0.5 to 2.0 for the $^2D_{3/2,5/2}$ states, whereas excellent agreements can be found for the 
	$^2P_{3/2,1/2}$ states. We have also noticed that the $\tau$ values for the $^2P_{1/2}$ state of the I$^{2+}$ ion, reported in Ref. 
	\cite{Biemont-AASS-1995} and the $\tau$ value for the $^2D_{3/2}$ state of the Po$^{+}$ ion, given in Ref. \cite{Biemont-PS-1996}, show very large 
	deviation from our results obtained using the MRCI method. The $\tau$ results for the Bi neutral atom and for the Po$^+$ and At$^{2+}$ ions provide 
	excellent agreement with the emperical results reported in Ref. \cite{Garstang-JRNBS-1964} and the HFR results reported in Ref. \cite{Biemont-PS-1996}.
	
	Our study reveals that there exist long-lived states, as can be seen from Table \ref{tau}, due to the fine-structure splitting of the $np^3$ 
	configurations such as for the $^2D_{3/2,5/2}$ states when $n$ is less than 6 and the $^2P_{1/2,3/2}$ states when $n$ is less than 4 of the under 
	investigated neural atoms and the singly and doubly charged ions. Transitions from these states to the ground state of the respective systems are
	almost in the optical regime. Therefore, these transitions can be very useful for making any high-precision measurements considering the resepective 
	atomic systems.
	
	\subsection{Hyperfine structure constants}
	
	Tables \ref{HFSA-1}, \ref{HFSA-2}, and \ref{HFSB-1} present the hyperfine structure constants $A$ and $B$ for the $^4S_{3/2}$, $^2D_{3/2,5/2}$, and
	$^2P_{1/2,3/2}$ states with the $n(=3-6)p^3$ configurations, which are obtained using the same computational procedure as that were used for the 
	evaluation of the $g_J$ and $\tau$ values. We compare our results with the experimental results whever available. We find that the Core18SV2SDV3SDT 
	model calculations used with the dyall.aae4z basis set give the a reasonably accurate predictions of the $A$ values. Further, we take into account 
	higher-order electron correlations by using the Core8SDV2SDTV3SDTQ model and increasing the cut-off of virtual orbitals from 10 a.u. to 20 a.u.. After all these efforts, our final results become 
	close to the available experimental values with deviations around 2-10 \%. The largest discrepancy is seen for the $^4S_{3/2}$ ground state, for which the 
	$A$ values are very sensitive to the inclusion of correlation effects from the inner-shell electrons and higher cut-off of the virtual orbitals. This 
	indicates that accuracy of the results can be improved further with the inclusion of correlation effects from higher level excitations and including 
	more number of virtual orbitals.  
	
	In contrast to the $A$ values discussed above, we do not find determination of the $B$ values using the DC Hamiltonian are very sensitive to inclusion of 
	correlations from the higher-level excitations or inclusion of high-lying virtual orbitals. Our results for the $5p^3$ $^2D_{3/2}$ and  $^2P_{1/2}$ states 
	of $^{121}$Sb are closer to their experimental values that are reported in Ref. \cite{Buchholz-ZPA-1978} but they deviate slightly from the values 
	reported in Ref. \cite{Hassini-JOSA-1988}. Similarly, our results for the $^2S_{3/2}$, $^2D_{3/2,5/2}$, and $^2P_{3/2}$ states $^{209}$Bi show excellent
	agreement with the available experimental values reported in Refs. \cite{George-JOSAB-1985,Wilman-JQSRT-2021,Wasowicz-PS-2007,Pearson-JPG-2000}.
	
	We have also calculated the $A$ and $B$ of the other singly and doubly charged ions with the $np^3$ configuration, for which no experimental results are 
	available. Based on our above analysis for the neutral atoms, we anticipate that accuracy of our calculated $A$ and $B$ values for these ions should be better 
	than 5-20\%. These predictions can be useful in identifying some of the astrophysical lines and in the plasma diagnosis processes. 
	
	\section{Conclusion}	
	
	We have determined energies, lifetimes, $g_J$ factors and hyperfine structure constants of the fine-structure splitting states having ground state
	$np^3$ configurations of neutral, singly, and doubly charged P-group atoms with principal quantum number $n=3-7$. For this purpose, we have employed 
	multi-configuration relativistic configuration interaction method using four-component Dirac-Coulomb Hamiltonian and relavistic Hamiltonian with 
	core-polarization effects. Uncertainties to the energies are obtained by extrapolating results using a complete set of basis functions and carry out 
	calculations by including higher level excitations to account for more electron correlation effects in both the form of approximated Hamiltonians. Wherever 
	available, we have compared our results for energies with the data listed in the NIST database and our predicted values for $\tau$, $g_J$ factor, and 
	hyperfine structure constants $A$ and $B$ with the previously reported experimental data, which show excellent agreements among them. This suggests that the
	reported results for which experimental values are not available are reliable enough to be useful various applications including in the analysis of 
	astrophysical and plasma diagnostic processes.

	\section{ACKNOWLEDGMENTS}
	
	This work is supported by The National Key Research and Development Program of China (2021YFA1402104), and Project supported by the Space 	
	Application System of China Manned Space Program. BKS would like to acknowledge use of ParamVikram-1000 HPC of Physical Research Laboratory (PRL), 
	Ahmedabad.

\end{document}